\documentclass[
aps,%
final,%
notitlepage,%
twocolumn,%
nobibnotes,%
nofootinbib,%
superscriptaddress,%
noshowpacs,%
centertags,
showkeys]%
{revtex4-2}

\usepackage{graphicx}
\usepackage{dcolumn}
\usepackage{bm}

\usepackage{amsmath,amssymb}
\usepackage[english]{babel}
\usepackage{graphicx}
\graphicspath{{./}{Figures/}}
\usepackage{xcolor}
\usepackage{cmap}
\usepackage{placeins}

\usepackage{hyperref}
\usepackage{soul}      
\usepackage{lipsum}    
\usepackage{blindtext} 
\usepackage{microtype} 

\definecolor{LightCyan}{rgb}{0.88,1,1}

\newcommand{\PWN}{\raise-0.4ex\hbox{\scalebox{0.8}{\scriptsize$P$\kern-0.05em$W$\kern-0.2em$N$}}}
\newcommand{\s}{\raise-0.1ex\hbox{\scalebox{1.2}{\scriptsize$s$}}}
\newcommand{\sla}{\;\raise0.55ex\hbox{\scriptsize$<$\kern-0.75em\raise-1.1ex\hbox{$\sim$}}\;}
\newcommand{\sga}{\;\raise0.55ex\hbox{\scriptsize$>$\kern-0.75em\raise-1.1ex\hbox{$\sim$}}\;}
\newcommand{\ssim}{\;\raise0.3ex\hbox{\tiny$\sim$}\,}
\newcommand{\sapprox}{\;\raise0.3ex\hbox{\tiny$\approx$}\,}

\def\lsim{\;\raise0.3ex\hbox{$<$\kern-0.75em\raise-1.1ex\hbox{$\sim$}}\;}
\def\gsim{\;\raise0.3ex\hbox{$>$\kern-0.75em\raise-1.1ex\hbox{$\sim$}}\;}

\def\kms{\rm ~km~s^{-1}}

\def\etal{{ et al. }}

\def \kms {\rm ~km~s$^{-1}$}

\def\ergs{\rm ~erg~s^{-1}}

\def\ecsb{erg cm$^{-2}$ s$^{-1}$ arcsec$^{-2}$ }
\def\ecsb2{erg cm$^{-2}$ s$^{-1}$ arcsec$^{-2}$}

\def\apj{ApJ}
\def\mnras{MNRAS}
\def\nat{Nat}

\def\araa{ARA\&A}                
\def\aap{A\&A}                   
\def\apjs{ApJS}                  
\def\apjl{ApJ}                   

\def\apss{Astroph. Space Sci.}

\def\ssr{Space Sci. Rev.}
\def\aapr{Astron. Astroph. Reviews}

\def\prl{Phys. Rev. Lett.}

\begin{document}

\title{Relativistic astrospheres of gamma-ray binaries:\\ modeling of non-thermal processes}

\author{A. M. Bykov}
\email{byk@astro.ioffe.ru}
\author{A. E. Petrov}%
 \email{a.e.petrov@mail.ioffe.ru}
 \author{K. P. Levenfish}
\affiliation{ 
Ioffe Institute, St.~Petersburg, 194021 Russia
}%

\begin{abstract}
A long standing problem in high energy astrophysics is the nature of galactic 
accelerators of particles with energies above petaelectronvolt (PeV). Such
objects are sources of galactic cosmic rays and  can produce PeV-regime 
photons observed by ground-based observatories. Among very likely 
accelerators are  astrospheres of pulsars in gamma-ray binaries. 
These binaries  have long been observed as bright sources of 
teraelectronvolt (TeV) gamma-ray photons. Recently, 2D relativistic 
magnetohydrodynamic (rMHD) simulations  have shown that the astrospheres can accelerate 
particles to energies well above PeV, provided  that they harbor a Gauss-range 
magnetic field. Such a strong field is necessary in the region of two colliding 
winds: the relativistic outflow of the pulsar or accreting black hole and the wind of its stellar companion, a massive early-type star. Here,  the wind collision region 
is explored as the site of PeV protons acceleration.
The local structure of colliding flows is illustrated using rMHD simulations 
of a powerful pulsar wind in 2D and 3D  models.
The relativistic outflow of a pulsar or black hole, evolving inside the strongly magnetized 
stellar wind, have an elongated shape and surrounded by a kind of magnetic cocoon providing 
favorable conditions for acceleration of ultra high energy ions. The simulated spectra of 
particles, accelerated by intermittent relativistic turbulence in these systems, have piece-wise 
power-law shape and extend well above PeV energies for powerful outflows.  
The model indicated that gamma-ray binaries harboring a powerful relativistic 
outflow, produced either by a pulsar or accreting black hole, can be   
bright sources of synchrotron MeV-regime photons and multi-PeV regime gamma-rays, as  
recently detected from galactic microquasars like Cyg X-3. The Gauss-range magnetic 
field  of a massive star wind strongly influences the non-thermal emission of gamma-ray binaries with relativistic companions.\\
\end{abstract}


\maketitle

\selectlanguage{english}

\section{Introduction}\footnote{This text is a slightly updated version of a review presented at  a conference ``Cosmic Gas Dynamics'' in memory of Professor V.B. Baranov dedicated to the 90th anniversary of his birth, published in Fluid Dynamics v.59, p. 2377, (2024)} \label{sec:intro}
Stellar winds as  gas and dust outflows accelerated in the upper atmosphere 
of stars of various types have long been observed as extended astrospheres 
\citep[see e.g.][]{Sobolev60, Baranov77, Lamers99, owocki23}. More than sixty years 
of studying the heliosphere with direct measurements of its key physical parameters 
provided unique information on the dynamics of the solar wind. Detailed models of 
the heliosphere have been created \citep{parker63,baranov71,Helio23}, taking into account 
the kinetics of its multiple components \citep[e.g.,][]{GodenkoIzmodenov23}. 
Powerful winds of young massive stars and compact relativistic objects, such 
as black holes and neutron stars, can also create extended astrospheres visible 
at many wavelengths.

The winds of neutron stars\,-\,pulsars power  energetic pulsar wind nebulae, 
known to be efficient accelerators of wind particles. The accelerated 
particles can convert up to a few tens of percent of the wind power into 
non-thermal synchrotron radiation from the nebula, ranging from radio to 
gamma-rays \citep{hester08}.  This turns pulsar astrospheres into bright 
synchrotron X-ray sources associated with sources of very high-energy 
gamma-rays.  
Such astrospheres can also produce energetic cosmic-ray (CR) leptons and are 
plausible antimatter factories that could explain the excess of CR positrons
observed in the GeV-TeV range at the Earth orbit \citep[e.g.,][]{Fang+19}. 
Under certain conditions, 
these objects can also accelerate CR nuclei.
Young pulsars with high spin-down luminosity $\dot{E}$, such as Crab pulsar, 
can accelerate particles even to PeV energies \citep{Arons12}. Synchrotron 
emission of PeV particles  falls in  the $\mbox{MeV}$--$\mbox{GeV}$ range 
if it occurs in magnetic fields of pulsar wind nebulae.
This study focuses on the astrospheres of rotation-powered pulsars  
as accelerators of very high-energy protons.

In binaries where a pulsar orbits a massive early-type star, the pulsar astrosphere may differ from that of an isolated pulsar. Such binaries are characterized by the collision of two flows, the relativistic wind of the pulsar and the powerful wind of its massive companion. The wind collision provides particularly favorable conditions for the efficient acceleration of both leptons and ions. The leptons can emit synchrotron radiation in the strong magnetic field of the system and inverse Compton radiation in the intense radiation field of the massive star \citep[see, e.g.][]{Dubus06b}. The hadrons, in turn, can emit gamma-rays by interacting with stellar wind nuclei and  with stellar optical/ultraviolet photons \citep[e.g.,][]{NeronovChernyakova07, NeronovRibordy09, 2032our}. The systems which appear bright in gamma-rays are called \textsl{gamma-ray binaries}; in general, they comprise a compact object (a neutron star  or a black hole) and a massive OB star.

The study of gamma-ray binaries is a challenging task, and a number of  
unsolved problems still present \citep{Dubus+13}. These include the 
unknown nature of the compact companion (black hole or pulsar?) in the gamma-ray 
binaries LS 5039 \citep{Yoneda+20,Kargaltsev+23,Makishima+23} and LSI $61^{\circ}303$ 
\citep{Weng+22, LoepezMiralles+23}, and the uncertainty 
over which component (leptons or hadrons?) is responsible for
the observed gamma-rays, to name a few.
Recent observations suggest that gamma-ray binaries might accelerate protons 
well above PeV. \textsl{LHAASO} gamma-ray observatory detected a 1.4 PeV photon from 
a source in the Cygnus region (which may be associated with the gamma-ray binary 
PSR J2032+4127) \citep{LHAASONat21}. A sub-PeV gamma-ray flare and neutrino were simultaneously
detected from the same region by \textsl{Carpet-2} and \textsl{Ice Cube} 
observatories \citep{Dzhappuev21}.

To solve the above problems it is necessary to  
model the non-thermal emission of gamma-ray binaries, taking into account the 
structure of flows in colliding winds. This structure appears to be quite complex, 
as follows from  hydrodynamic (HD) and magnetohydrodynamic (MHD) simulations
\citep[see, e.g.][]{BoschRamonBarkov11,BoschRamon+12,BoschRamon+15,Dubus+15,BoschRamon+17,Huber+21}. 
Most of 
the above simulations neglected the effects  of the magnetic field of the stellar wind. 
However, these effects can be important for particle acceleration, 
especially in  short-period or highly eccentric gamma-ray binaries. 

To explain the observed sub-PeV  
flares 
and neutrinos as produced by protons, 
the latter must be accelerated above PeV. This requires the presence of 
a Gauss-range magnetic field in the region of wind collision, as 
shown by Monte Carlo simulations \citep{2032our}. 
Such a strong field can be carried by the wind of O/B/A-type stars, since 
about 10\% of them are found to have surface fields above $100\:$G  
\citep[][]{OBstars_Bfield17,Bstar_Bfield19}. 
Indeed, the known orbital parameters of observed gamma-ray binaries imply that
the orbital separation of their compact and massive companions could range 
from $\sim 0.1$ AU to tens of AU, i.e. from units to hundreds of typical radii of  
massive stars, $R_{\star} \sim 10^{12}\:$cm.
Assuming the Parker-type model for the radial structure of the stellar wind \textit{(sw),}\ 
with $B_{sw} \propto R_{\star}/ R$, one would expect $B_{sw} \ssim 0.1$--$10\:$G  
in the vicinity of the compact object.

The study of gamma-ray binaries as potential pevatrons is of great interest now.
Monte Carlo modeling \citep{2032our} of these systems was performed under 
simplified assumptions on the structure of their flows.
A more rigorous consideration requires combining relativistic MHD and particle-in-cell (PIC) approaches 
to describe the dynamics of colliding flows and the propagation of high-energy 
particles in them, respectively. 
Such simulations were carried out  by \citet{Bykov+24} in a planar geometry with a limitet dynamical range.    
They studied  the local structure of colliding flow in the vicinity 
of a pulsar, taking into account the strong magnetic field of the stellar wind.
A strong field is necessary to keep the accelerating particles inside the accelerator 
during the time interval they gain energy above PeV. This time interval was shown to be short compared to 
the orbital period, which justified the local consideration. 
Whether a planar geometry is applicable for such a task was a question for study.
In this paper we  compare results of the planar and full 3D rMHD simulations 
of gamma-ray binaries that harbor a pulsar. 
The simulation provided a detailed description of the structure of relativistic flows in these objects and
highlight the effect of a strong magnetic field of the stellar wind. We consider the model results in a broad context of the study of non-thermal processes in relativistic astrospheres produced by either young rotation powered pulsar or  black hole(or neutron star) accreting at a critical rate.
\section{MHD MODEL SETUP} \label{sec:model}
 Self-consistent MHD-PIC modeling of gamma-ray binaries requires resolving
 their MHD structures on spatial scales smaller than the  particle gyroradii,  
 $R_g$.  For a proton of energy $\sim 0.1$ PeV in a one-Gauss field, the estimate is
 $R_g \sim 0.02\:$AU.  Meanwhile, the region of colliding flows 
 of  interest to us can cover tens of AU.
The numerical grid should then have $\ssim 1000$ nodes  per dimension 
to ensure the minimum necessary dynamic range of spatial scales, 
even if particle acceleration is tracked starting at sub-PeV energies.
Starting at lower energies would entail a much higher resolution.
 The need  for a fine grid is compounded by  the need  to scan the parameter 
 space of the models so that they can provide a realistic description of 
 non-thermal emission of  gamma-ray binaries.
 Such high computational demands on the problem motivate us to look for 
 a more cost-effective approach.

A hint about the first simplification follows from the short time scales 
of the physical processes under consideration. Bykov et al.\ (2024) \citep{Bykov+24} found 
that it takes significantly less than $10^5\;$s for the complete expansion 
of the nebula and the accompanying acceleration of protons to PeV energies, 
provided that the expansion starts from zero in a stellar wind with a 
Gauss-range field. This is very short compared to the orbital period of 
known gamma-ray binaries with pulsars \citep{Casares05,Ho+17}\footnote{Interestingly, gamma-ray binaries with black holes that have very short periods have also smaller orbital separation distance and, therefore, stronger magnetic field in the colliding flow region, that, in pricinple, allows similar consideration}. This is even shorter 
than the estimated periastron passage time during which one would expect 
efficient acceleration of PeV regime protons in the case of elongated orbits with periods more than a year like in PSR B1259-63 and PSR J2032+4127. Rapid inflation allows 
the nebula to easily adjust its flow structures to local stellar wind conditions.
Moreover, according to \citep{Bykov+24}, the strong magnetic field in the wind collision region becomes the main factor shaping its structure, making the magnetic force dominant over the Coriolis force induced by the orbital motion of the pulsar.
With this in  mind and taking into account the rapid acceleration of particles,
one can consider only the \textsl{local structure}\ of the region of 
the colliding flows. This greatly simplifies the problem. There is little point in 
monitoring this region for most of the pulsar's orbit, where the field's influence on the structure is weak or the parameters of the wind collision zone do not facilitate multi-PeV particles acceleration.

In the considered objects, the region of direct wind collision can be very compact. 
According to \citep{Bykov+24}, the nebula bubble can occupy a volume 
more than
two orders of magnitude smaller than it would have inside the supernova 
remnant. If this bubble is small compared to the distance between the binary companions,
the problem can be simplified even further (especially given the shortness 
of the dynamical and acceleration time scales). In the \textsl{local}\ vicinity 
of the bubble, the stellar wind flow can be considered  uniform, with fixed 
directions of field and velocity. Another simplification follows from the fact 
that in a strongly  magnetized stellar wind, the flow velocity tends to align with 
the magnetic field.
    
The above reasoning reduces a computationally efficient setup for studying the
acceleration of PeV protons in gamma-ray binaries to the following. 
We consider a pulsar wind nebula inflating in a dense, uniform, strongly-magnetized 
flow with aligned magnetic field and velocity.
The magnitudes of the field and velocities, as well as pressure and mass density, 
are prescribed by modern  stellar wind models \citep[e.g.,][]{Bogovalov+08,Klement+17}.
The consideration is local and includes the nebula itself and its immediate surroundings.
\citet{Bykov+24}  noted that the acceleration of multi-PeV particles may not 
be sensitive to the symmetry of the system: only the size of the acceleration 
site and the spectrum of upscattering magnetic inhomogeneities in it 
play a decisive role. 
With this in mind, they performed  planar (2D) modeling of gamma-ray binaries. 
Here we simulate and analyze the flows structure in 2D and 3D and discuss the 
validity of planar models.

\section{MODELING RESULTS} \label{sec:results}
\subsection{2D and 3D nebulae developing in a strongly magnetized stellar wind} \label{ssec:2D_3D_vs_B}
\begin{figure*}
\begin{minipage}{1.0\textwidth}
\begin{minipage}{.29\textwidth}
        \includegraphics[width=\textwidth]{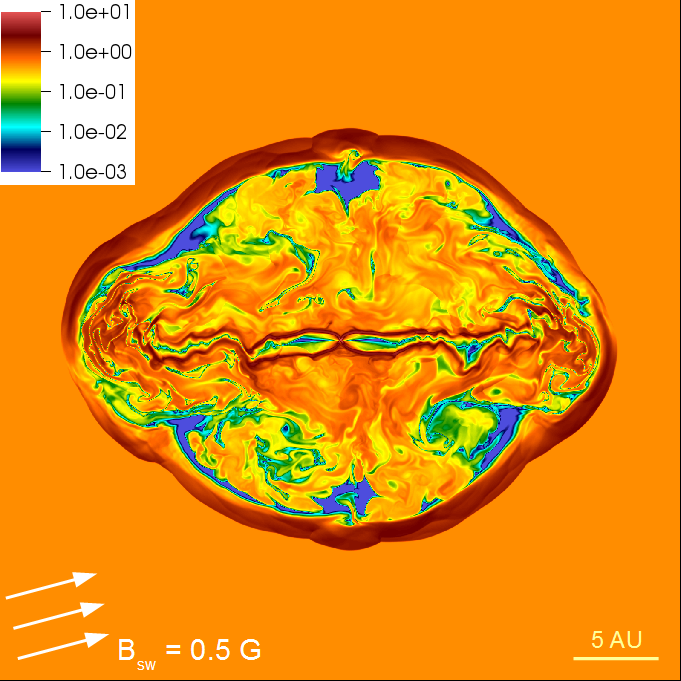} 
   \end{minipage}
      \nolinebreak
   \hfill
\begin{minipage}{.29\textwidth}
        \includegraphics[width=\textwidth]{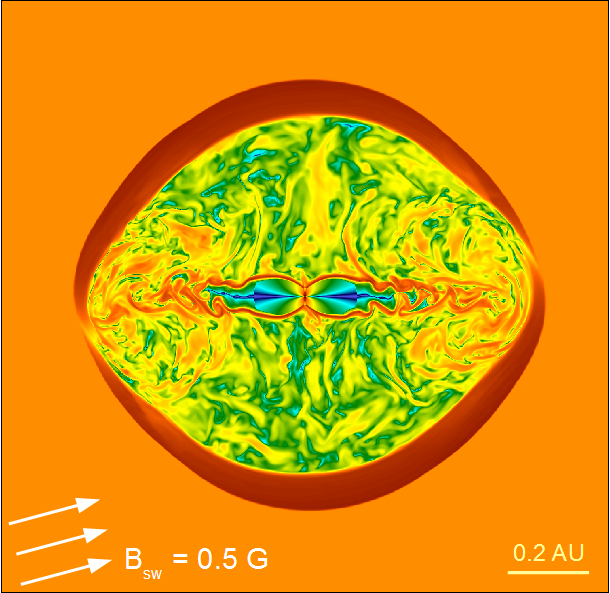} 
   \end{minipage}
   \nolinebreak
   \hfill
   \begin{minipage}{.29\textwidth}
        \includegraphics[width=\textwidth]{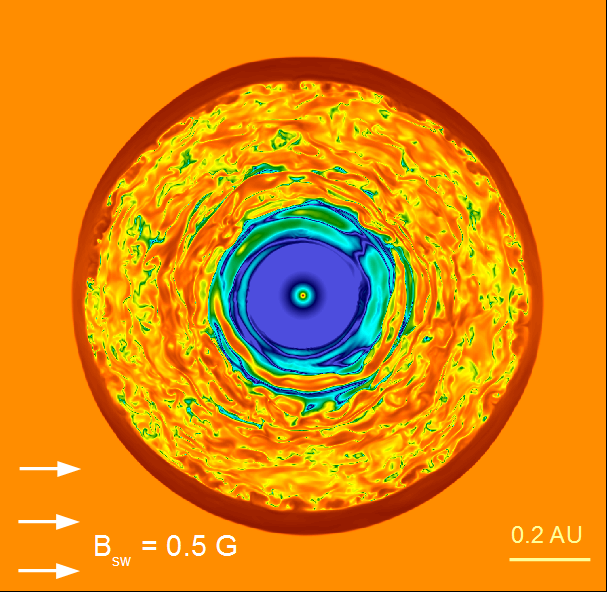}
   \end{minipage}
      \linebreak
   \vfill
   \begin{minipage}{.29\textwidth}
        \includegraphics[width=\textwidth]{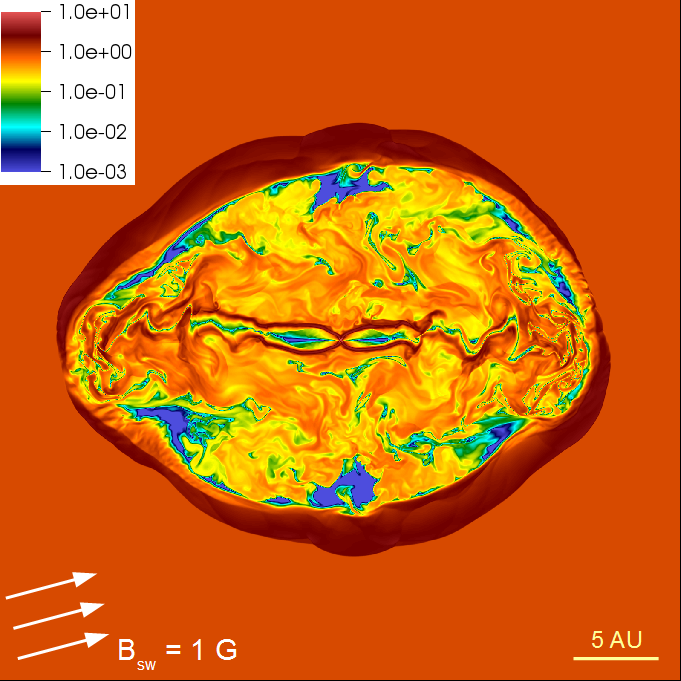}
   \end{minipage}
   \nolinebreak
   \hfill
   \begin{minipage}{.29\textwidth}
     \includegraphics[width=\textwidth]{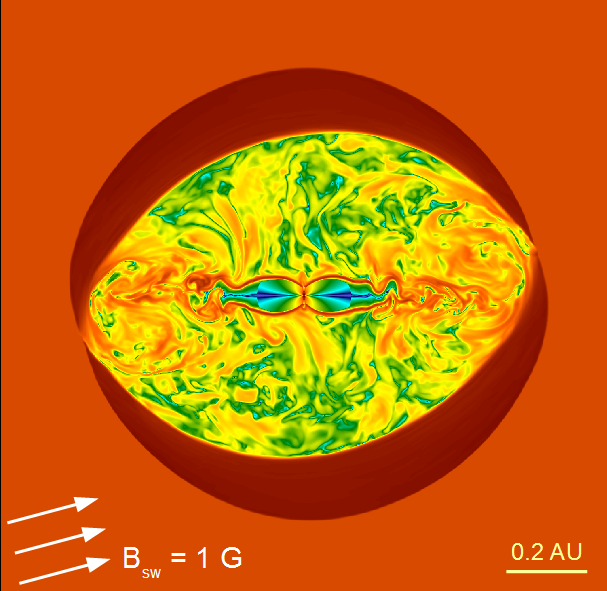}
   \end{minipage}
   \nolinebreak
   \hfill
   \begin{minipage}{.29\textwidth}
     \includegraphics[width=\textwidth]{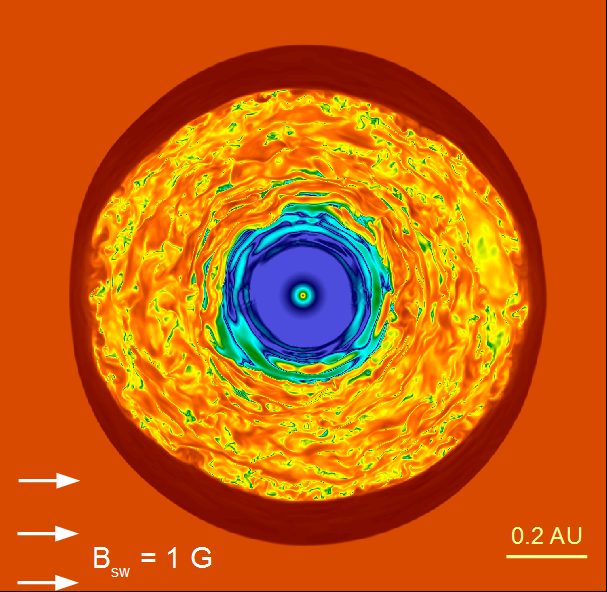}
   \end{minipage}
   \linebreak
   \vfill
   \begin{minipage}{.29\textwidth}
        \includegraphics[width=\textwidth]{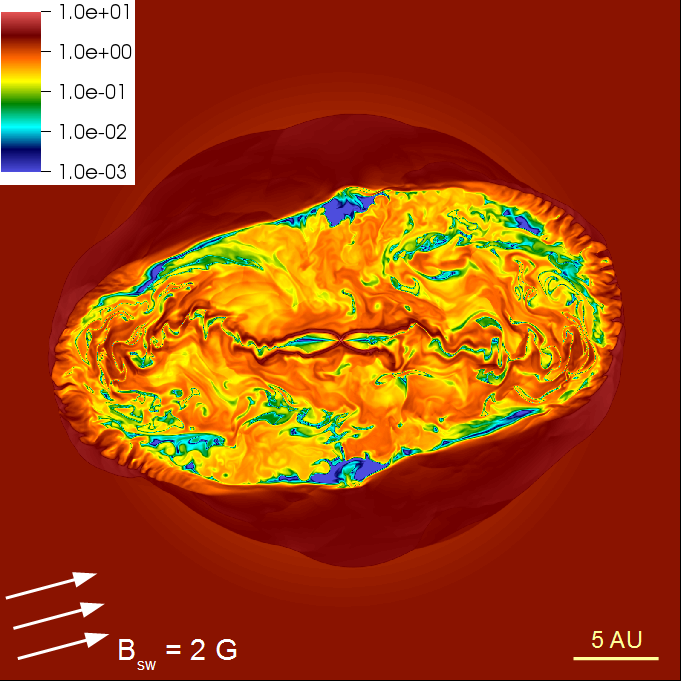}
   \end{minipage}
   \nolinebreak
   \hfill
   \begin{minipage}{.29\textwidth}
     \includegraphics[width=\textwidth]{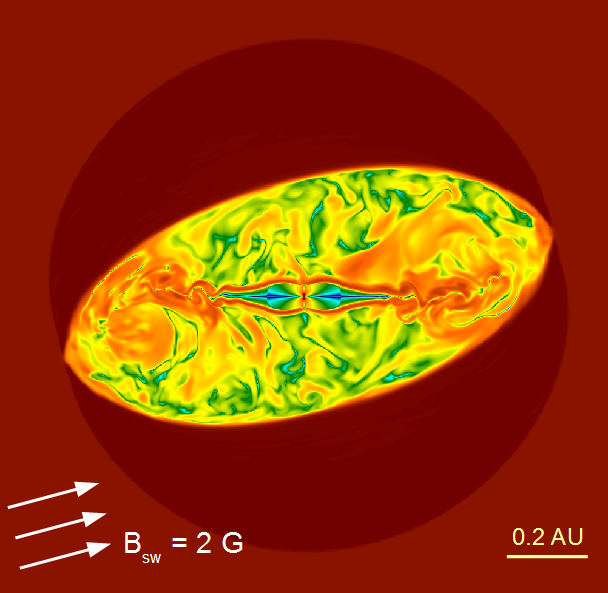}
   \end{minipage}
   \nolinebreak
   \hfill
   \begin{minipage}{.3\textwidth}
     \includegraphics[width=\textwidth]{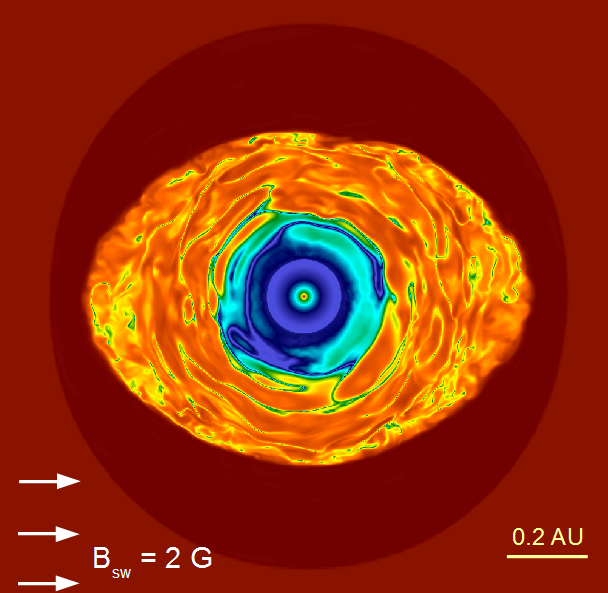}
   \end{minipage}
       \linebreak
   \vfill
   \begin{minipage}{.29\textwidth}
        \includegraphics[width=\textwidth]{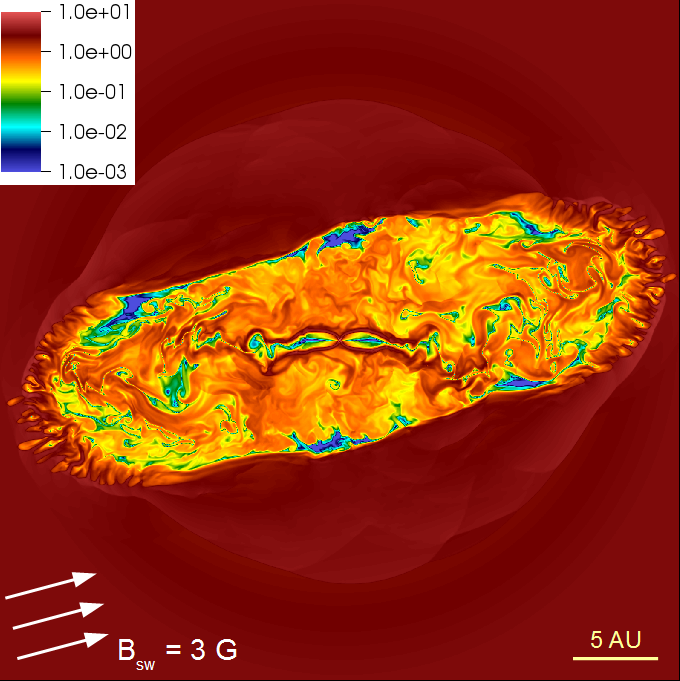}
   \end{minipage}
   \nolinebreak
   \hfill
   \begin{minipage}{.29\textwidth}
     \includegraphics[width=\textwidth]{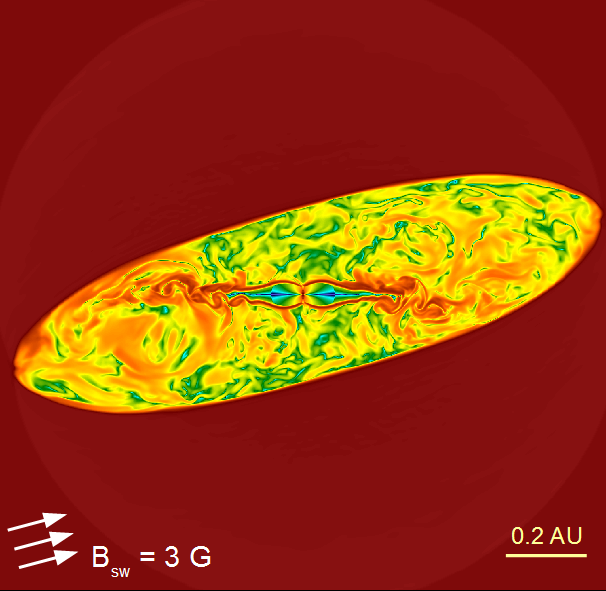}
   \end{minipage}
   \nolinebreak
   \hfill
   \begin{minipage}{.29\textwidth}
     \includegraphics[width=\textwidth]{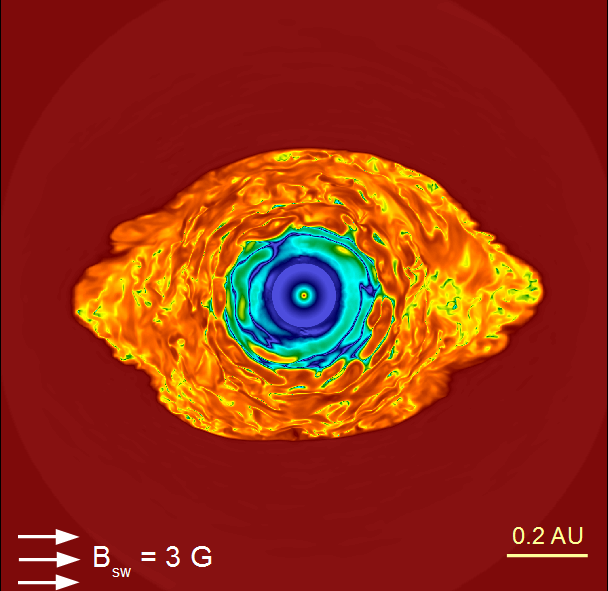}
   \end{minipage}
 \caption{\label{PWN_vs_B_2D_and_3D}%
 Wind collision region in 2D and 3D runs. Colors on the maps indicate 
 the local magnetic field $B$ [G]. 
 \underline{Left column:\vphantom{g}}\ 2D model (meridional cut). 
 \underline{Middle and right columns:}\ 3D model (meridional 
 and equatorial cuts).  
The rows illustrate the effect
of increasing the strength of the stellar wind field $B_{sw}$.
From top to bottom:  $B_{sw} = $ 0.5, 1, 2 and 3 Gauss. 
$\vec{B}_{sw}$ is shown by white arrows and makes  
 $ 75^{\circ}$  with the pulsar's rotation axis  (pointing upward in 
 the 1st and 2nd columns and  out-of-plane in the 3rd one).
The color palette (the same in all panels) highlights the flow structures, 
not the maximum and minimum values of $B$.  2D and 3D setups corresponds to models 
\textsl{A1}--\textsl{A4} and \textsl{X1}--\textsl{X4} in Table \ref{Table_2D_3D};  
we show their snapshots taken 15 and 2 hours after initiation, respectively.}
   \end{minipage}
 \end{figure*}

To compare the 2D and 3D structures of a nebula immersed in a highly magnetized 
stellar wind  \textit{(sw)}, we ran two corresponding sets of rMHD models and tested
them for different  
values of $B_{sw}$.
The models are listed in Table \ref{Table_2D_3D}, and Fig.\ \ref{PWN_vs_B_2D_and_3D} depicts the results.
For any of the studied models, the nebula forms within a few hours. 
Its MHD structure, arising  from interactions with the stellar wind, 
then evolves in a self-similar manner.
The structure is shown in Fig. \ \ref{PWN_vs_B_2D_and_3D} 
using the $B$ field maps. The maps are snapshots of a 2D nebula in its
meridional cut (left column) and a 3D nebula in its meridional and 
equatorial cuts (middle and right columns, respectively).
In each column, the maps are arranged to illustrate the effect
of increasing the strength of the stellar wind field $B_{sw}$.

In the meridional cuts,  the structure of magnetized flows in 2D and 3D runs
turns out to be very similar and can be  briefly described as follows.
The pulsar, sitting in the center of the bubble, emits cold relativistic 
pulsar wind \textit{(blue region).} The wind then abruptly decelerates 
and heats up at the termination shock. The wind zone and shock both appear 
$\infty$-shaped 
because the wind is stronger  near 
the nebula's equator than near its poles \citep{Michel73}. 
Not only the power but also the wind field is anisotropic: it reaches a 
maximum at mid-latitudes and drops sharply toward the poles and equator 
(e.g., \citep{Lyubarsky02}; see details in Appendix \ref{sec:Appenix-A-rMHD}). Because 
of the combination of these anisotropies, the wind 
flow approaching the shock is weakly magnetized at equatorial latitudes 
and strongly magnetized at higher latitudes, where the shock is closer to 
the pulsar. When crossing the inclined  parts of the shock wave, the 
 wind's radial streamlines are deflected and plasma magnetization increases 
-- again, depending on latitude. A smooth change in the angle of deflection  
along the inclined (arch-shaped) parts of the shock allows the latter to collect 
strongly magnetized plasma into two narrow-channel  streams   \textit{(shown 
in dark red),} tightly pressed against its front, and focus them toward the 
equator. There, the streams begin to mix and intertwine. Because they originate 
in hemispheres with different magnetic polarities, their mixing results in 
a turbulent, highly magnetized outflow. Its magnetic inhomogeneities can propagate 
outward with $v/c \sim 0.5$--$0.85$, depending on the model. To a distant observer, 
such outflow could result in the appearance of an X-ray torus  (or double-torus), 
a synchrotron feature often seen in synchrotron images of subsonic or slightly 
supersonic pulsar wind nebulae (PWNe) 
\citep{Weisskopf+00,Hester+02,Pavlov+01,Helfand+01,Pavlov+03,hester08,Kargaltsev+15}.

In  many ways the picture described above resembles that seen in models of 
isolated pulsar wind nebulae. Examples include models of the subsonic 
Crab Nebula \citep{KL04,delZanna+04,Camus+09,Porth+14,Olmi+14,Olmi+16,BuhlerGiomi16,Levenfish+21}
and the slightly-supersonic Vela nebula \citep{Ponomaryov+23}.
However, there are a number of differences, which are listed below.
(1) In gamma-ray binaries, the nebulae  are much 
more compact, therefore they accumulate the magnetic energy much faster.
Their magnetic fields are in the sub-Gauss range and can even reach $\ssim$Gauss, 
while in isolated nebulae they barely exceed a few hundred micro-Gauss. 
 (2) Another difference is that an X-ray torus may look tilted toward the equator 
of the nebula.   
The tilt of the torus can gradually disappear as the pulsar approaches 
apastron and recover again as it approaches periastron.
(3)  In gamma-binaries, the nebula bubble is surrounded by a dense magnetic cocoon 
that forms from the stellar wind material perturbed by the bubble’s expansion. 
This cocoon is of paramount importance for accelerating very high-energy particles,
and we will discuss it a little later.
(4) The nebula can form relativistic clumps -- unusually-fast magnetic inhomogeneities 
not observed (at least not yet) in models of isolated nebulae. 
These clumps can act as scattering centers, capable of providing 
particularly efficient acceleration of high-energy particles.
(5) Finally,  the nebular bubble can be significantly contracted 
across the field direction, as it described below.

Along the $B_{sw}$ field direction the size of the bubble is determined 
by the balance of the ram pressure of the stellar wind and the pressure 
inside the nebula. 
Across this field, the bubble expansion is additionally limited by the 
tension of the magnetic field lines. The way in which  the bubble proportions 
respond to increasing  $B_{sw}$ is illustrated by the successive rows in Fig.\  
\ref{PWN_vs_B_2D_and_3D} in which this field  is taken to be  
$\:0.5, \:1, \:2$ and $\: 3$ G, from top to  bottom. 
Comparing the maps in the left and middle columns, one can see that the 
2D and 3D bubbles evolve in a similar way and have  similar shapes, 
despite the strong differences in their pulsar and stellar wind parameters  
(see Table \ref{Table_2D_3D}). This is especially noticeable 
in the limit of a strong field of 2 to 3 G. 
 The shape of the bubble is formed under the influence of two anisotropies: 
the latitudinal anisotropy of the pulsar wind power and the one caused  
by  the misalignment of $\textbf{B}_{sw}$ 
and the pulsar's rotation axis (which 
is vertical in the plane of the maps).
At lower fields -- 0.5 and 1 G -- the first factor predominates, and 
the bubble is more extended in the equatorial direction, around which 
the pulsar wind is more powerful. At stronger fields -- 2 and 3 G --
the magnetic tension of the $\textbf{B}_{sw}$ lines  takes over, and the bubble 
stretches along their direction. The field strength at which this transition 
occurs can be estimated by equating the magnetic pressure $B_{sw}^2/8 \pi$ 
in the stellar wind to the momentum density flux at a distance $r$ 
from the pulsar, $\Phi = \dot{E} / 4 \pi r^2 u_{pwn}$, 
carried  by nebular \textit{(pwn)} flows expanding at a velocity $u_{pwn}\:$:
\begin{equation}
\begin{aligned}
B_{sw} & = \left(\frac{2 \dot{E}}{u_{pwn}\: r^2}\right)^{1/2} \\ & =0.5 \:\dot{E}_{37}^{1/2} \:\left(\,\frac{u_{pwn}}{0.1 c}\,\right)^{-1/2}\: \left(\,\frac{r}{10\: \mbox{AU}}\, \right)^{-1} \;\mbox{G}\, .
\end{aligned}
\label{Elongated_PWN_B_estimate}
\end{equation}
The estimate suggests that in fields $B_{sw}\,\ssim 1\:$G 
 and above the nebula bubble will be noticeably elongated in the directions of 
the field, even if the wind power is as high as $\dot{E}\sim 10^{37}\:$ erg$\,\,$s$^{-1}$.
This is indeed observed in Fig.\ \ref{PWN_vs_B_2D_and_3D}, both in 2D and 3D runs. 
For a given $B_{sw}$, the ratio of the longitudinal to transverse size 
for 2D and 3D bubbles is similar: it is $\simeq 0.5$ for $2\:$G and $\:\simeq 0.3$--$0.35$ 
for $3\:$G.

At the same time, the equatorial size of 2D and 3D bubbles (on meridional 
cuts) remains practically unchanged, and therefore the size of the X-ray torus too,
despite the fact that the bubble shrinks in the total volume.    
The torus is simply reshaped from axially 
symmetric to oval due to a slight flattening across the field when it is 
almplified by $\sim 6$ times. In addition, the torus gradually acquires 
a tilt toward the nebula's equator (determined by the plane of the pulsar's 
rotational equator). This occurs because the 
strongly magnetized equatorial outflow, which creates the appearance of 
the torus to a distant observer, tends to be deflected in different directions 
from the equator on the leeward and windward sides of the nebula. In 
essence, 
this outflow  gradually transforms from disk-shaped to S-shaped with 
increasing field strength in the stellar wind.
The tilt of the torus can 
change gradually  along the pulsar's orbit,
reaching a minimum at apastron and a maximum at periastron.

Although at kpc distances the modern X-ray telescopes 
cannot resolve X-ray tori smaller than 100 AU in size, 
their radio counterparts  may, in principle, be resolvable by radio telescopes.
The change in the tilt of the torus, assuming that the effect is 
observed, can be used to track the change in the $B_{sw}$ strength along 
the pulsar's orbit. The effect can even help to constraint the strength 
itself, if other parameters affecting the torus appearance have already 
been constrained at the apastron (assuming that the field effects
are negligible there). In sub-Gauss fields, the X-ray torus will be highlighted 
by synchrotron radiation from electrons with Lorentz facors of about $10^6$.
Such electrons can, in principle, come directly with the pulsar wind, so their 
number
can be few orders of magnitude greater than that of accelerated electrons with 
Lorentz factors above $10^8$, which are usually responsible for the soft X-ray 
radiation of an isolated pulsar's nebula.

Unlike the torus, the shock wave retreats as the field increases. 
This is most clearly seen in the right column in Fig.\ \ref{PWN_vs_B_2D_and_3D}. 
This behavior is expected. The dynamic pressure of the stellar wind
increases as $\rho u^2_{sw} + B_{sw}^2 / 8 \pi$, causing the bubble 
to decrease in volume. 
So the average magnetization of the plasma in the remaining volume goes up.
This brings into play an effect well known in modeling nebulae of isolated 
pulsars: as the average magnetization in the nebula increases, the termination 
shock moves closer to the pulsar, and the polar jets\footnote{
    In our models, jets -- strongly collimated polar outflows --  
    are not observed, since their formation requires the use of a much finer numerical grid than ours; see, e.g., \cite{Porth+14}, 
    for more details.
} become more powerful \cite{KL04, delZanna+04}.
Our simulations indicate that the local MHD structure of the pulsar wind nebula 
depends more than moderately on the velocity and density of the stellar wind, 
as long as the latter is strongly magnetized.

\begin{figure*}
\begin{minipage}{1\textwidth}
 \begin{minipage}{.24\textwidth}
 \center{\includegraphics[width=1.0\textwidth]{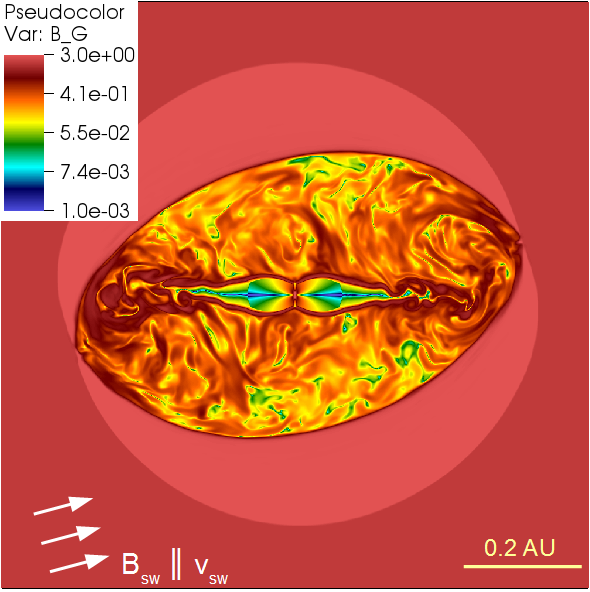}}
\end{minipage}
\nolinebreak
\hfill
\begin{minipage}{.31\textwidth}
 \center{\includegraphics[width=1.0\textwidth]{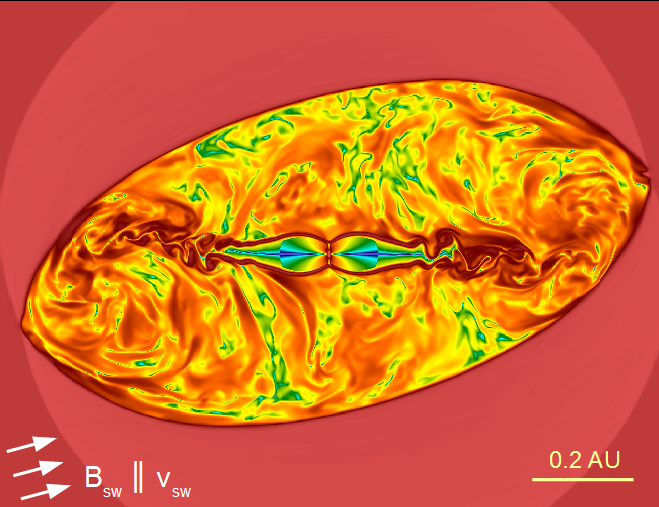}}
\end{minipage}
\nolinebreak
\hfill
\begin{minipage}{.448\textwidth}
{ \includegraphics[width=1.0\textwidth]{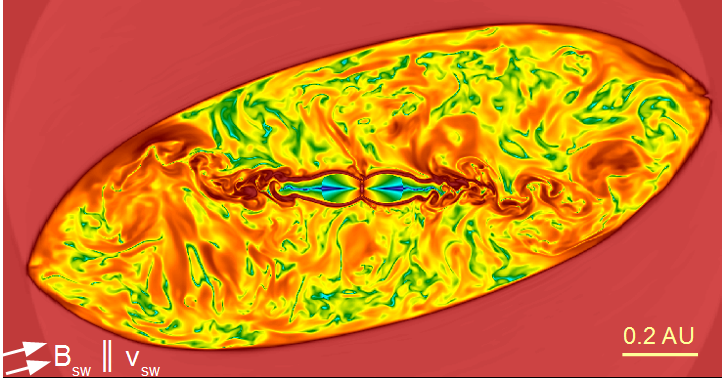} }
\end{minipage}
\linebreak
\vfill
\begin{minipage}{.24\textwidth}
 \center{\includegraphics[width=1.0\textwidth]{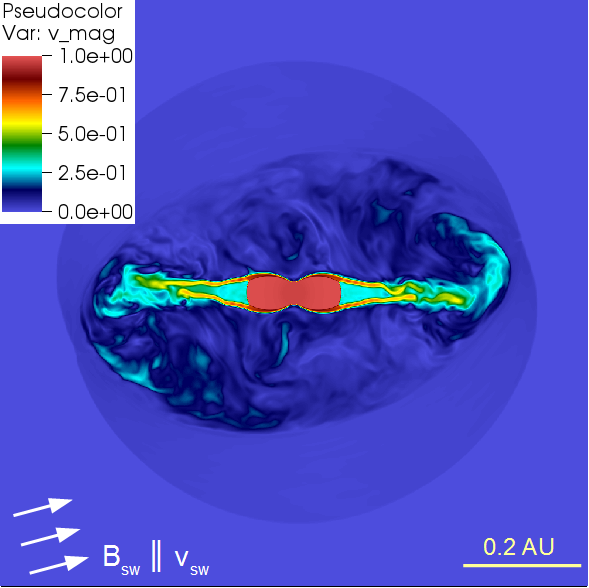}}
\end{minipage}
\nolinebreak
\hfill
\begin{minipage}{.31\textwidth}
 \center{\includegraphics[width=1.0\textwidth]{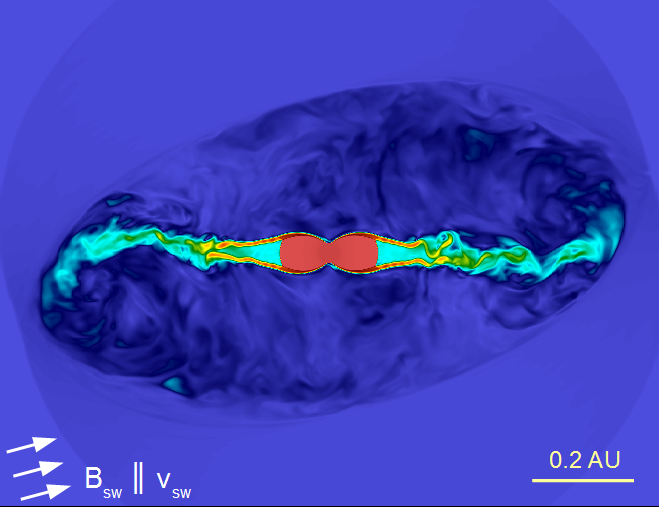}}
\end{minipage}
\nolinebreak
\hfill
\begin{minipage}{.448\textwidth}
{ \includegraphics[width=1.0\textwidth]{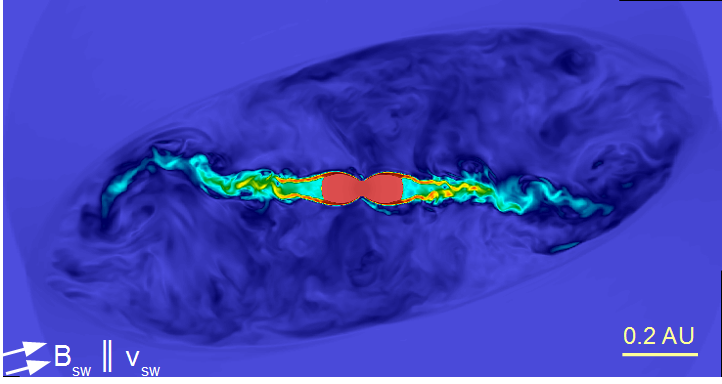} }
\end{minipage}
\linebreak
\vfill
\begin{minipage}{.24\textwidth}
 \center{\includegraphics[width=1.0\textwidth]{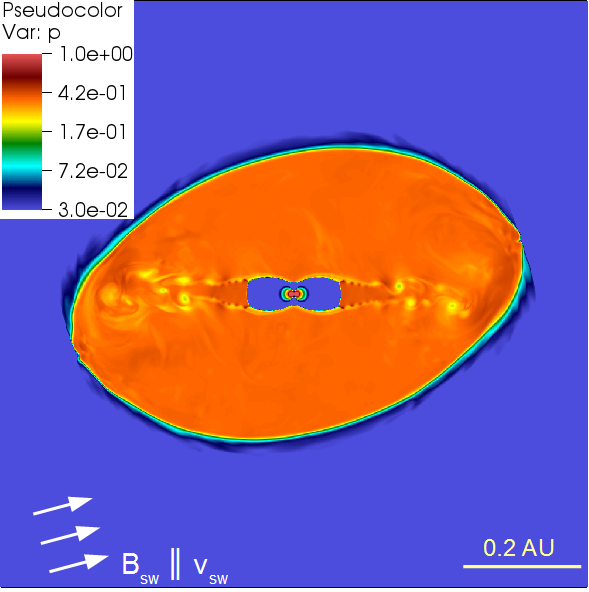}}
\end{minipage}
\nolinebreak
\hfill
\begin{minipage}{.31\textwidth}
 \center{\includegraphics[width=1.0\textwidth]{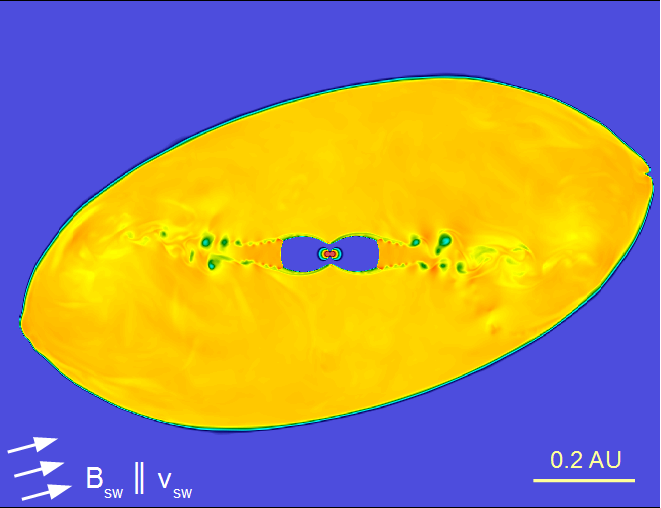}}
\end{minipage}
\nolinebreak
\hfill
\begin{minipage}{.448\textwidth}
{ \includegraphics[width=1.0\textwidth]{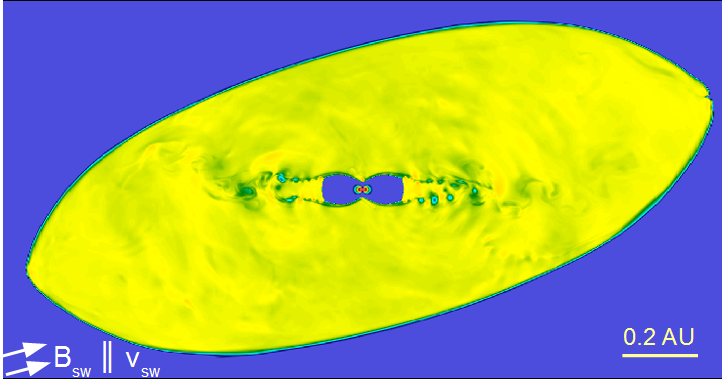} }
\end{minipage}
\end{minipage}
\caption{Time evolution of 
the pulsar wind nebula in gamma-ray binaries (3D case).  The 3D nebula 
is shown in section by the plane defined by the pulsar's rotation axis 
and the magnetic field vector of the stellar wind. The rows from top 
to bottom show maps of the magnetic field (in Gauss), velocity (in $c$ 
units), and pressure (in cgs units). The maps describe the MHD flows (as 
they appear 
in the \textit{X3} model from Table \ref{Table_2D_3D} in which $B_{sw}=2\:$G). 
In each row, three 
consecutive maps are obtained at 0.8, 2, and 3 (model) hours after the 
onset of nebula inflation. The brightness scale of the color bar (universal 
for all maps in a particular row) is adjusted to highlight the flow patterns 
and does not reflect the maximum and minimum values of $B$, $v$ and $p$.
}
\label{3D_temp_evol_B_2}
\end{figure*}

Let us turn to  the magnetic cocoon.
The expanding bubble perturbs the stellar wind, which eventually leads  
to the creation of a shell of compressed stellar-wind matter around the 
nebula.
This  shell can be several times denser than the undisturbed wind. 
The field in the shell (being frozen into the plasma, $B \propto \rho$)
can also be several times stronger than the unperturbed 
$B_{sw}$, and fall 
into the Gauss-range even if $B_{sw}$ is at sub-Gauss. In essence, 
the shell acts as a kind of magnetic cocoon capable of holding high energy 
particles within the accelerator -- the wind collision region that encloses 
the nebula and the cocoon itself. The PeV particles may only be held there 
for a few hours, but 
that is still long enough  for them to gain an energy above 10 
PeV before leaving the region \citep{Bykov+24}. As $B_{sw}$ increases, 
the bubble gives off more and more of its volume to the magnetic cocoon that 
forms around it. At the same time, the total volume of the region of wind 
collision changes little, at least in the 2D case (as can be seen in the left 
column in Fig.\ \ref{PWN_vs_B_2D_and_3D}). 
In the 3D case the total volume seems  to increase, but this is most likely 
a numerical effect.
Due to their resource-intensive nature, our 3D runs 
were not as long as the 2D runs; they were stopped when we were convinced 
that the 3D nebula had reached the stage of self-similar evolution (Fig.\ 
\ref{3D_temp_evol_B_2}); 
it seems that this was not enough time for the cocoon to fully form.
So the effect is still to be checked. Finally, we note the presence of 
noticeable magnetic and density inhomogeneities in the cocoon. They are 
barely visible in Fig.\ \ref{PWN_vs_B_2D_and_3D} since its color 
palette is aimed at highlighting the MHD structures in the body of the 
nebula, not in the cocoon. Inhomogeneities form despite the fact that 
in our models of gamma-ray binaries the unperturbed wind of a massive 
companion is considered to be homogeneous, with a uniform magnetic field.  
The dynamics of these inhomogeneities is still to be studied, but their 
very presence may be important for the confinement  and acceleration of 
high-energy particles.

As shown above, the 2D and 3D results have much in common. However, 
it is worth noting some of their differences. For the same gamma-ray 
binary parameters, the pressure inside the three-dimensional nebula 
is more uniform, and the MHD flows are less turbulent. These findings 
are  consistent with the results of \citet{Porth+14}, pioneers in 3D 
modeling of isolated pulsar wind nebulae. They showed that the third degree 
of freedom destroys  the artificially increased coherence of MHD flows 
in the presence of a purely toroidal field in the (axisymmetric) 2D nebula 
and softens their impact on the termination shock. Being more stable, the 3D shock in turn  excites less energetic driven-scale turbulence in the downstream \cite{Camus+09,Ponomaryov+23}. Overall, 
a more uniform pressure in the nebula leads to a smoother shock wave profile; it also smooths out the outlines of the cocoon and nebula itself, making  
the cocoon's irregularities less pronounced.\\

\subsection{Relativistic clumps}\label{ssec:clumps}
In their 2D models of gamma-ray binary systems,  \citet{Bykov+24} 
noted the appearance in 
the nebular outflows of
“relativistic clumps” --
especially fast-moving magnetic inhomogeneities with Lorentz factors 
$\varGamma \gsim 3$ and sub-AU sizes.
Such clumps are of great importance for Fermi-type particle acceleration 
to energies above PeV. Accelerating particles gain energy 
in head-on collisions with magnetic inhomogeneities in turbulent flows. 
In a single head-on collision, a particle can boost its energy by a factor 
of $\varGamma^{\,2}$, i.e., by the squared Lorentz factor of the encountered 
inhomogeneity. For the energy gain to be so significant, the inhomogeneity 
must be comparable in size  to the particle's gyroradius. The sub-AU 
scales of the relativistic clumps allow them to effectively upscatter 
pre-accelerated sub-PeV particles, since the latter have gyroradii $\sla 1\:$AU 
in the Gauss-range magnetic fields typical of nebulae in gamma-ray binaries. 
For the clumps, the factor $\varGamma^{\,2}$ is almost an order of magnitude 
as large as that for the usual magnetic inhomogeneities. Therefore, the clumps 
are much more efficient and can turn a sub-PeV particle into a PeV one in just 
a few successful scatterings.

The nature of the large-scale relativistic clumps remains to be elucidated. 
The clumps seem to form differently than the usual large-scale magnetic inhomogeneities. 
The latter arise in the form of magnetic eddies 
at the rims of the wind termination shock due to strong distortions 
of its working surface\footnote{
	Note that modern relativistic MHD models of PWNe are built on a simplified 
	prescription for the pulsar wind, in which the wind itself is uniform 
	(in terms of density); see details in Appendix A.}
 \citep{Camus+09,Porth+14}.
The shock wave then throws them into the nebula, where they quickly cascade
and are advected outward by the turbulent equatorial flow. Their  $\varGamma$-factors 
can be estimated from observations of the Crab Nebula and from rMHD simulations of 
nebulae in general. In Crab, the usual inhomogeneities are seen in radio, optical, 
and X-rays as fine, prominently bright arch-shaped synchrotron features (the 
so-called \textit{wisps}\,). They 
originate at the wind termination shock and propagate outward with 
$v/c\sim 0.5$--$0.6$, corresponding to $\varGamma \sim 1.2\;$ \citep{Hester+02}. 
Similar Lorentz factors are  demonstrated  by large-scale eddies in the equatorial 
flows in rMHD models\footnote{
    This estimate applies to flows that have already been thrown into the nebula 
    from the arched vaults of the termination shock. Flows still running along 
    these vaults are very narrow and quasi-laminar and therefore  
    cannot upscatter a PeV-regime  particle, no matter how high their $\varGamma$ is
    (this $\varGamma$ is estimated in \cite{KomissarovLyutikov11}).
    }
of Crab-like nebulae, objects with a single-torus morphology in X-rays.
A characteristic feature of these objects is that flows of opposite 
magnetic polarities, arising in different hemispheres of the nebula, 
meet and mix right after 
the wind termination shock, thereby giving rise to a turbulent equatorial flow 
(details on the topic can be found in recent reviews \cite{Porth17,Amato20,OlmiBucciantini23}).

Meanwhile, the relativistic clumps appear to originate not at the 
shock, but at some distance from it, deep in the nebula. They certainly 
correlate with strong  depressions that occur on either side of the wide, 
overcompressed, weakly-magnetized equatorial flow, where it temporally 
collapses or narrows greatly. 
Such equatorial flow is characteristic of another type  of pulsar wind 
nebulae represented by Vela, an object with a double-torus X-ray morphology 
(Sect.\ \ref{ssec:Crab-and-Vela}). To form a double-torus, a pulsar must be 
in sub-sonic or slightly-supersonic motion relative to the external 
matter and have a wind with low average magnetization ($\lsim 0.002$;  
see  \cite{Ponomaryov+21,Ponomaryov+23} for more details).  Models 
\textsl{X1}--\textsl{X4} in Table \ref{Table_2D_3D} meet these conditions, 
so their flow pattern resembles that in the Vela-type nebulae. 
There,  outflows of opposite polarity on either side of the 
overcompressed equatorial flow remain strongly-magnetized, quasi-laminar 
and ultra-relativistic, and do not mix up to distances of $2$--$3$ shock radii. 
Their typical Lorentz factors can come to  $\varGamma \sim 1.5$--$1.9$  
(corresponding to $v/c \sim 0.7$--$0.85$) \cite{Ponomaryov+21,Ponomaryov+23}.
 The clumps seem to arise from  the interplay of these strongly-magnetized 
relativistic flows with the overcompressed weakly-magnetized flow at 
the nebula's equator.

\begin{figure*}
\begin{minipage}{\textwidth}
\begin{minipage}{0.47\textwidth}
\includegraphics[width=\textwidth]{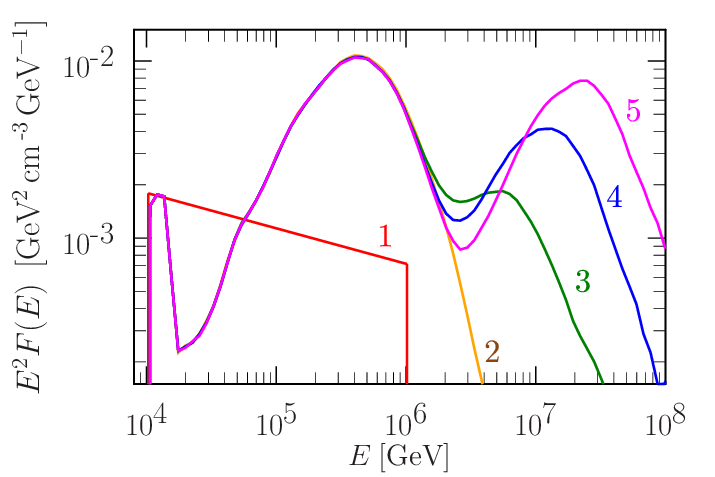}
\end{minipage}
\nolinebreak
\hfill
\begin{minipage}{0.52\textwidth}
\includegraphics[width=\textwidth]{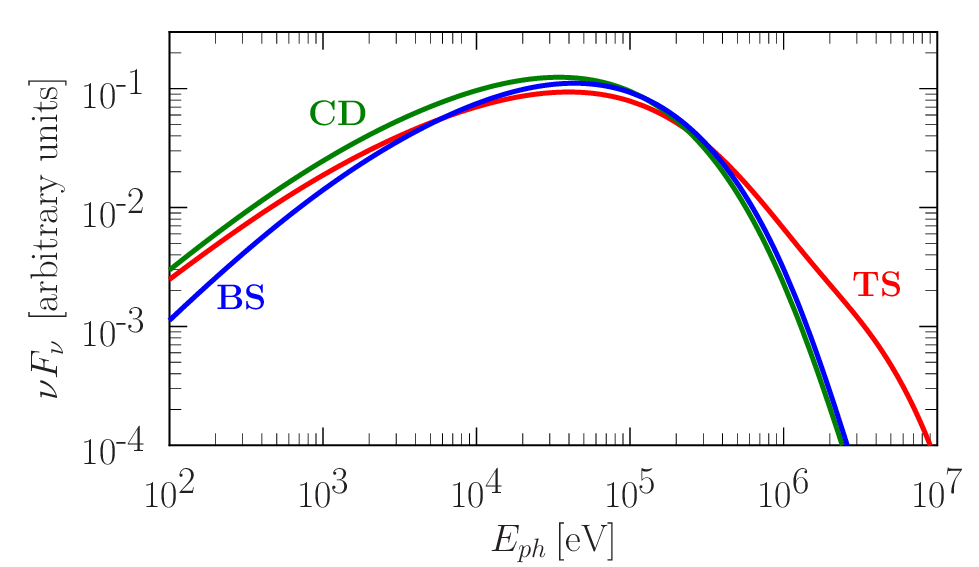}
\end{minipage}
 \end{minipage}
\caption{\underline{Left:}  Particle acceleration in the wind collision region of gamma-ray binaries (results of 
Monte-Carlo simulations). Several spectral energy distributions are shown in color: 
in red (1) -- for the particles  just injected in the system, 
in orange (2) -- for those that were accelerated in the colliding flows alone, 
and in green, blue and magenta (3,\,4,\,5) -- for those that were accelerated in 
the colliding flows with the relativistic clump. The last three spectra coinciding  with the orange curve below $1 \:\mbox{PeV} = 10^{6} \:\mbox{GeV}$ differ 
in the clump’s Lorentz factor, which is $\varGamma = 3$, 4.5 and 6, respectively.
\underline{Right:} simulated synchrotron spectra of the Vela-like pulsar wind nebula.
Different colors refer to the synchrotron emissivity integrated 
over different lines of sight. Namely, over lines of sight coming close: 
to the termination shock \textit{(TS, in red)}, to the bow shock \textit{(BS, in blue)},  and
to the contact discontinuity \textit{(CD, in green)}.
}
\label{MC_spectra}
\end{figure*}

The role of relativistic clumps in particle re-acceleration can be illustrated 
using a simple Monte-Carlo model (Appendix \ref{sec:Appendix-B-anis-MC}). In it, a clump is modeled as 
a subregion embedded in the region of the relativistic wind, which in turn collides 
with the stellar wind.  The embedded subregion is stationary, has a sub-AU size 
and high velocity with a certain $\varGamma$. A population of pre-accelerated 
particles is then injected in the system and allowed to interact with it. The 
result can be seen in Fig.\ \ref{MC_spectra} on the left panel.
It shows several spectral energy distributions 
of the injected particles; namely, those  that were:\\
  \hspace*{.5em}(1) \parbox[t]{.9\linewidth}{just injected into the system;}\\
  \hspace*{.5em}(2) \parbox[t]{.9\linewidth}{accelerated in colliding flows alone;}\\
  \hspace*{.5em}(3/4/5)\:\parbox[t]{.9\linewidth}{accelerated in colliding flows with the clump\\ 
        (with Lorentz factors  $\varGamma=$3/4.5/6, respectively).}
It can be seen that the acceleration in the colliding flows alone makes the 
spectral distribution much harder, with a single hump at sub-PeV energies. 
Embedding a clump with $\varGamma \sga 3$ in the colliding flows results in a 
double-humped spectrum with a second hump  at energies well above PeV. 

\begin{figure*}
\begin{minipage}{1.0\textwidth}
\begin{minipage}{.041\textwidth}
        \includegraphics[width=\textwidth]{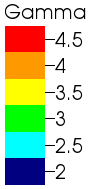} 
   \end{minipage}
   \nolinebreak
   \hfill
\begin{minipage}{.46\textwidth}
        \includegraphics[width=\textwidth]{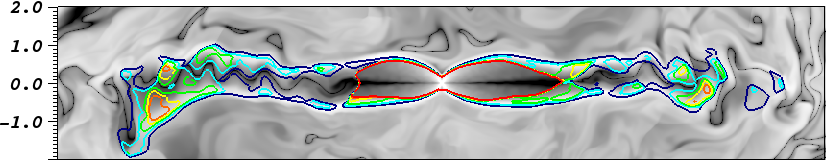} 
   \end{minipage}
   \nolinebreak
   \hfill
   \begin{minipage}{.435\textwidth}
        \includegraphics[width=\textwidth]{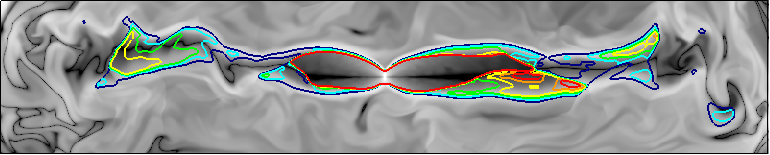}
   \end{minipage}
   \nolinebreak
   \hfill
   \begin{minipage}{.063\textwidth}
        \includegraphics[width=\textwidth]{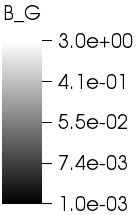} 
   \end{minipage}
   \linebreak
   \vfill
 \begin{minipage}{.041\textwidth}
        \includegraphics[width=\textwidth]{colorbar_Gamma_2_4p5.png} 
   \end{minipage}
   \nolinebreak
   \hfill
\begin{minipage}{.46\textwidth}
        \includegraphics[width=\textwidth]{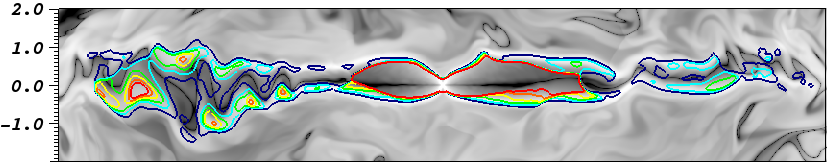} 
   \end{minipage}
   \nolinebreak
   \hfill
   \begin{minipage}{.435\textwidth}
        \includegraphics[width=\textwidth]{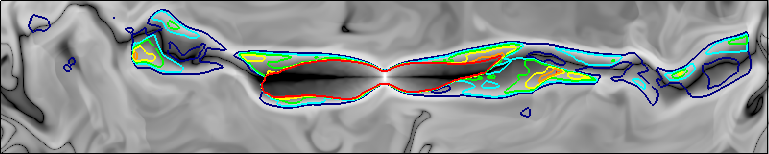}
   \end{minipage}
   \nolinebreak
   \hfill
   \begin{minipage}{.063\textwidth}
        \includegraphics[width=\textwidth]{colorbar_B_gray.png} 
   \end{minipage}
   \linebreak
   \vfill
   \begin{minipage}{.041\textwidth}
        \includegraphics[width=\textwidth]{colorbar_Gamma_2_4p5.png} 
   \end{minipage}
   \nolinebreak
   \hfill
\begin{minipage}{.46\textwidth}
        \includegraphics[width=\textwidth]{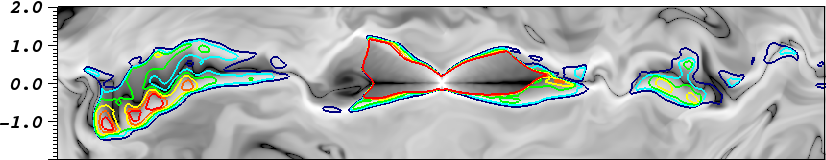} 
   \end{minipage}
   \nolinebreak
   \hfill
   \begin{minipage}{.435\textwidth}
        \includegraphics[width=\textwidth]{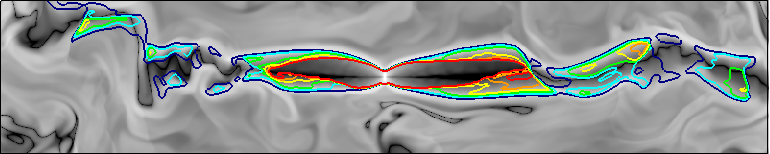}
   \end{minipage}
   \nolinebreak
   \hfill
   \begin{minipage}{.063\textwidth}
        \includegraphics[width=\textwidth]{colorbar_B_gray.png} 
   \end{minipage}
   \linebreak
   \vfill
   \begin{minipage}{.041\textwidth}
        \includegraphics[width=\textwidth]{colorbar_Gamma_2_4p5.png} 
   \end{minipage}
   \nolinebreak
   \hfill
\begin{minipage}{.46\textwidth}
        \includegraphics[width=\textwidth]{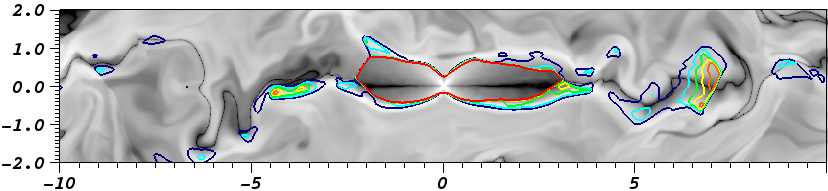} 
   \end{minipage}
   \nolinebreak
   \hfill
   \begin{minipage}{.435\textwidth}
        \includegraphics[width=\textwidth]{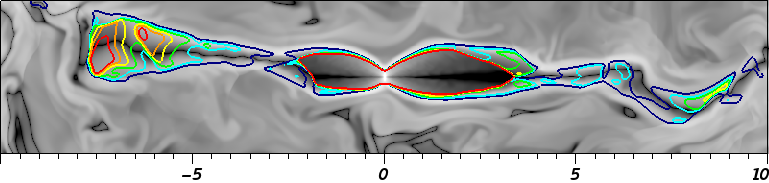}
   \end{minipage}
   \nolinebreak
   \hfill
   \begin{minipage}{.063\textwidth}
        \includegraphics[width=\textwidth]{colorbar_B_gray.png} 
   \end{minipage}
   \linebreak
   \vfill
 \caption{\label{Clumps_on_B_2D}Relativistic clumps in pulsar wind nebulae (enlarged views of  the 
 equatorial region). 
 Shown are  2D models \textit{A5}\ and \textit{A6}\ in Table \ref{Table_2D_3D}: 
 $\alpha = 45^{\circ}$, $\sigma_0 = 0.3$ (left column) and $\alpha = 80^{\circ}$, 
 $\sigma_0 = 3$ (right column). Contour plots of the Lorentz factor are superimposed 
 onto  gray-color maps of the magnetic field $B$ [in G] provided to guide the eye.
 Each row corresponds to a certain point in time since the start of simulation: 
 $t = 11.1$, $11.8$, $12.5$ and $13.2$ hours, from top to bottom. 
 The gray-color bar is adjusted to highlight the structure of rMHD outflows and 
 does not reflect the maximum and minimum values of $B$. Note that the contour 
 $\varGamma_{pw}=4.5$, that outlines the region of the unshocked pulsar wind, almost 
 coincides with the position of the wind termination shock, due to the abrupt 
 deceleration of the wind. The axes units are AU.}
  \end{minipage}
 \end{figure*}

According to our 2D simulations, the clumps  can arise in  PWNe 
with very different parameters (see Fig. \ref{Clumps_on_B_2D}). 
The clumps with $\varGamma \gsim 3$ can be more than 1 AU in size,  
i.e. larger, than the gyroradius of $10$ PeV particles in a $1 \:$G  
field. These clumps can persist in  turbulent  flows of the nebula 
for a few hours, which is more than enough time for  re-acceleration to occur 
\citep{Bykov+24}.

\begin{figure*}
\begin{minipage}{1.0\textwidth}
\begin{minipage}{.06\textwidth}
        \includegraphics[width=\textwidth]{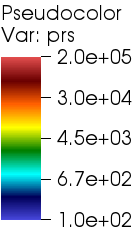} 
   \end{minipage}
\nolinebreak
\hfill
\begin{minipage}{.43\textwidth}
        \includegraphics[width=\textwidth]{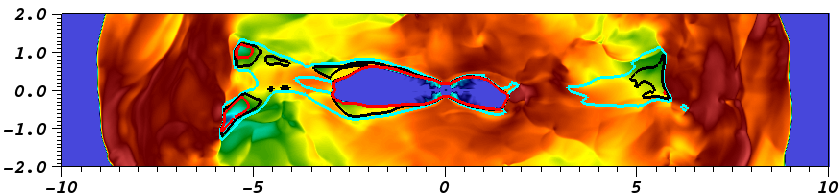} 
   \end{minipage}
   \nolinebreak
   \hfill
   \begin{minipage}{.43\textwidth}
        \includegraphics[width=\textwidth]{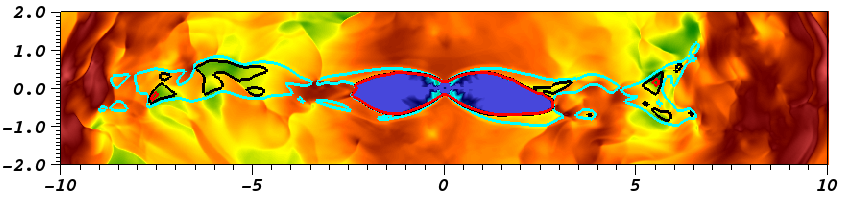}
   \end{minipage}
   \nolinebreak
   \hfill
   \begin{minipage}{.06\textwidth}
        \includegraphics[width=\textwidth]{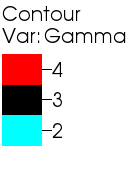} 
   \end{minipage}
   \linebreak
   \vfill
   \begin{minipage}{.06\textwidth}
        \includegraphics[width=\textwidth]{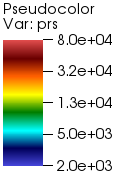} 
   \end{minipage}
   \nolinebreak
   \hfill
   \begin{minipage}{.31\textwidth}
        \includegraphics[width=\textwidth]{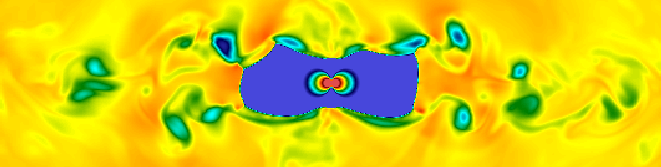} 
   \end{minipage}
   \nolinebreak
   \hfill
   \begin{minipage}{.31\textwidth}
        \includegraphics[width=\textwidth]{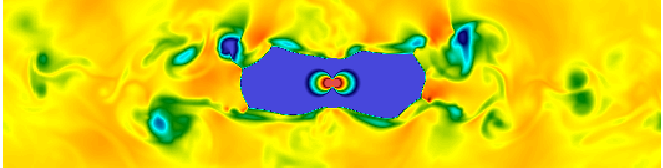} 
   \end{minipage}
   \nolinebreak
   \hfill
   \begin{minipage}{.31\textwidth}
        \includegraphics[width=\textwidth]{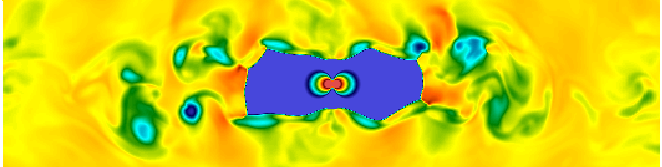} 
   \end{minipage}
   \linebreak
   \vfill
   \begin{minipage}{.06\textwidth}
        \includegraphics[width=\textwidth]{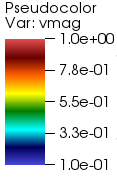} 
   \end{minipage}
   \nolinebreak
   \hfill
   \begin{minipage}{.31\textwidth}
        \includegraphics[width=\textwidth]{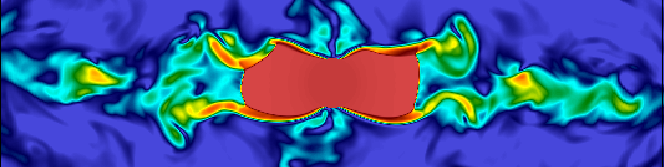} 
   \end{minipage}
   \nolinebreak
   \hfill
   \begin{minipage}{.31\textwidth}
        \includegraphics[width=\textwidth]{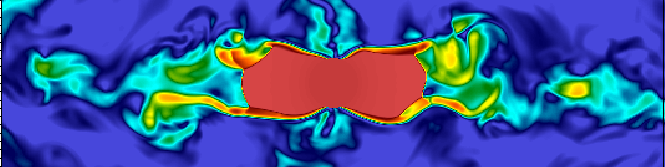} 
   \end{minipage}
   \nolinebreak
   \hfill
   \begin{minipage}{.31\textwidth}
        \includegraphics[width=\textwidth]{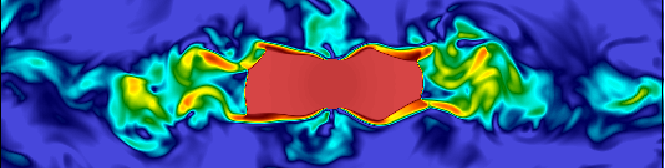} 
   \end{minipage}
 \caption{\label{Clumps_on_P} 
 Relativistic clumps in pulsar wind nebulae  (enlarged views of  the 
 equatorial region).  Results of 2D/3D  rMHD  simulations  of the 
 colliding winds region in gamma-ray  binaries.  \underline{Top row 
 (2D case)}:  Colored pressure maps with superimposed  contour plots 
 of Lorentz factors 
 of MHD-structures 
 with $\varGamma>2$.  The maps  shows the 
 nebula $7.6$ and $9.7\:$ (model) hours after the onset of its inflation. 
 The calculation is  based on the  model \textit{A5}\  from Table 
 \ref{Table_2D_3D} ($\alpha = 45^{\circ}$ and $\sigma_0 = 0.3$). The axes units are AU.
 \underline{Middle} \underline{and bottom rows (3D case):} Correlation of depressions and clumps 
 with a high Lorentz factor. Shown are colored maps of pressure (middle row) 
 and flow velocity (in units of $ c$; bottom row). 
 Successive maps in the rows are obtained at time points spaced $\sim 100$  
 seconds apart, starting at 1.3 (model) hours after the onset of nebula inflation.
 The calculation is based on the model \textit{X6}\ in Table \ref{Table_2D_3D} 
 ($\alpha = 60^{\circ}$, $\sigma_0 = 0.1$).   
 Pressure colorbars are in  units of $p_0 = 9\cdot 10^{-6}\:$ dyn$\cdot$cm$^{-2}$.
 The brightness scales of  the colorbars are adjusted to highlight the structure  
 of flows and   do not reflect the maximum and minimum  values of the quantities.}
  \end{minipage}
 \end{figure*}

\begin{figure}
\begin{minipage}{\linewidth}
   \begin{minipage}{.26\textwidth}
        \includegraphics[width=\textwidth]{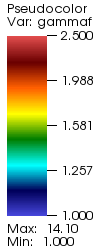} 
   \end{minipage}
   \nolinebreak
   \begin{minipage}{.73\textwidth}
        \includegraphics[width=\textwidth]{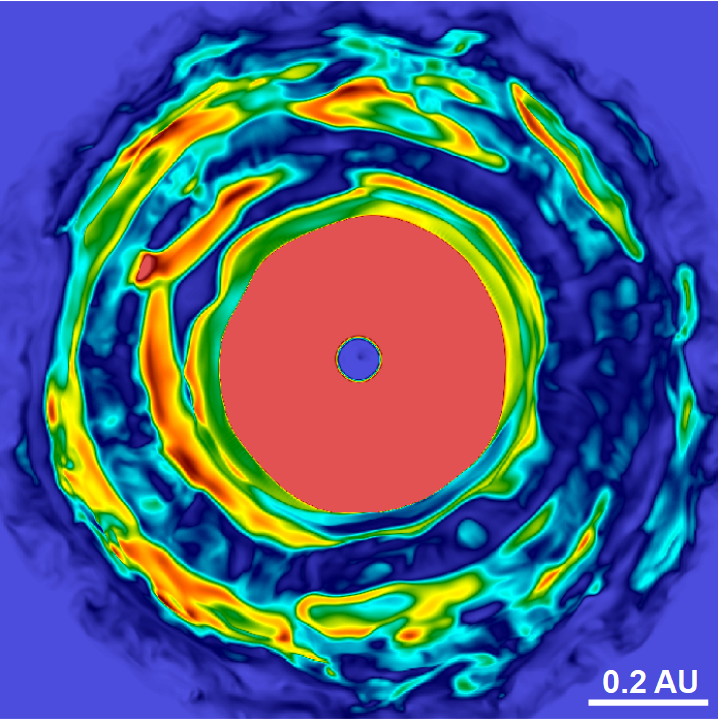} 
   \end{minipage}
\end{minipage}
 \caption{\label{Clumps_gamma_map_a80s03_3D}Color map of the  Lorentz-factor $\varGamma$  of rMHD flows 
 (3D model nebula X5 from Table \ref{Table_2D_3D}). Shown is a cross-section of the nebula by a plane 
 parallel to its equatorial plane and offset from it by 0.058 AU. The brightness scale of the color bar 
 is adjusted to highlight the structure of rMHD flows and does not reflect the maximum and minimum 
 values of $\varGamma$.}
 \end{figure}

 \begin{figure}
\begin{minipage}{1.0\linewidth}
\begin{minipage}{0.8\linewidth}
        \includegraphics[width=\linewidth]{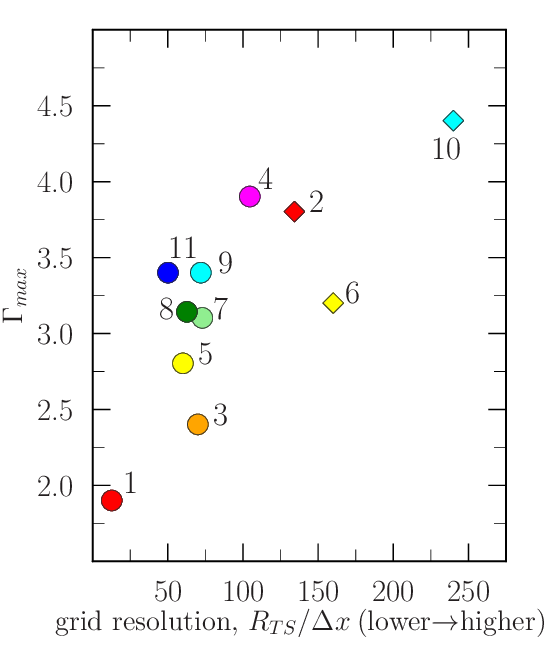} 
   \end{minipage}
   \linebreak
   \vfill
   \begin{minipage}{\linewidth}
        \begin{tabular}{lccccc}\hline
    Model & $\alpha$ & $\sigma_0$  & $n_{sw},\:\mbox{cm}^{-3}$ & $B_{sw}$,\:G & $\psi$,$^{\circ}$\\ \hline 
    1 & 45$^{\circ}$ & 0.03 & 300 & 0.1 & 90 \\
    2 & 45$^{\circ}$ & 0.03 & 300 & 0.1 & 90 \\
    3 & 45$^{\circ}$ & 0.1 & 300 & 0.1 & 90 \\
    4 & 45$^{\circ}$ & 1 & 300 & 0.1 & 90 \\
    5 & 85$^{\circ}$ & 0.03 & 300 & 0.1 & 90 \\
    6 & 85$^{\circ}$ & 0.03 & 300 & 0.1 & 90 \\
    7 & 45$^{\circ}$ & 0.03 & 3000 & 0.4 & 45 \\
    8 & 45$^{\circ}$ & 0.03 & 30000 & 0.4 & 45 \\
    9 & 45$^{\circ}$ & 0.03 & 30000 & 0.1 & 45 \\
    10 & 45$^{\circ}$ & 0.03 & 30000 & 0.1 & 45 \\
    11 & 45$^{\circ}$ & 10 & 30000 & 0.4 & 45 \\
    \hline
\end{tabular}
   \end{minipage}
 \caption{\label{Clumps_gamma_vs_resolution}
 Maximum Lorentz factor of relativistic clumps as a function of numerical grid resolution, 
 for different 2D rMHD setups (shown in the table). 
 The resolution is parameterized by the number of grid's nodes
 within the equatorial radius $R_{TS}$ of the termination shock. 
 Symbols of the same color correspond to models with the same setup, 
 which differ only in grid resolution. 
}
  \end{minipage}
 \end{figure}

In agreement with the result of \citet{Porth+14}, in our 3D nebulae
the termination shock is less dynamic and magnetic turbulence is less 
intense. Despite this, we found that particularly fast-moving magnetic 
clumps still form in the nebula,  although smaller in size and with 
lower $\varGamma$. This can be seen in Fig.\ \ref{Clumps_on_P}, in the 
bottom two rows. In models with $\alpha = 80^{\circ}$ the clumps have 
$\varGamma \gsim 2$ and sizes of several percents of AU in the 
meridional plane (where the pulsar's spin axis lies) and several 
tenths of AU in the plane parallel to the equatorial plane of the 
nebula (Fig.\ \ref{Clumps_gamma_map_a80s03_3D}). Whether this decrease 
in size and speed of the clump is a feature of 3D nebulae or a consequence 
of the insufficiently high resolution of 3D models remains to be seen. 
Our calculations indicate that the 3D clumps are larger, faster and more 
numerous the more strongly the pulsar wind is magnetized.

Relativistic clumps appear over a wide range of grid resolutions 
and over a large region of parameter space. As expected, their sizes 
and Lorentz factors depend on resolution of the numerical 
grid\footnote{
     they also depend on the convergence of the numerical scheme on 
     which the numerical code is based;  examples in the field of PWNe 
     modeling can be found, e.g.,  in \citet{Porth+14}
     }. This is illustrated in  Fig.\ \ref{Clumps_gamma_vs_resolution}  
for a number of 2D setups. This is a standard situation within the framework 
of an  ideal rMHD approach,  in which artificial numerical viscosity (determined 
by grid's resolution) 
substitutes the real plasma viscosity. 

\subsection{Astrospheres of isolated pulsars: the case of Crab and Vela}
\label{ssec:Crab-and-Vela}
Like nebulae in gamma-ray binaries, nebulae of isolated pulsars can also be 
subject to a strong external flow. Examples include the Crab and Vela X-ray
nebulae, not to mention highly supersonic objects. Crab appears to be influenced 
by the ejecta flow caused by  the asymmetric expansion of the supernova remnant.
This strong flow moves in almost the same direction as the pulsar, but faster 
than it. Taking this fact into consideration is important  for interpreting 
synchrotron features observed in the  nebula -- its asymmetric jets, and the 
\textit{``inner ring''} associated with the termination shock \cite{Levenfish+21}.
In turn,  Vela encounters a fast counter-flow that appears slightly  supersonic 
in the pulsar's reference frame; it was initiated by a reverse shock of a 
supernova that swept over the nebula some time ago \citep{ChevalierReynolds11}.  
Taking this counter-flow into account is necessary for interpreting virtually 
all synchrotron features of Vela's X-ray morphology,  from its enigmatic 
double-torus to its complexly structured asymmetric jets \citep{Ponomaryov+21, Ponomaryov+23, Fateeva+23}. 
The counter-flow is also the most natural cause of the quasi-periodic dynamics of these  features, including the observed reverberations of the double-torus and even its precession \cite{Petrov+23},  signs of which were detected by the Chandra X-ray telescope \cite{Kargaltsev+02,Durant+13}.

In the presence of a strong  external flow, the nebular flow pattern 
may become more regular, and the wind termination 
shock more stable\footnote{this refers to the stability of the 
shock geometry, not the shock transition itself}. Earlier  2.5D rMHD 
simulations\footnote{
 2.5D refers to an axisymmetric numerical grid without equatorial symmetry
} \cite{Ponomaryov+21} 
showed that the shock stability is an essential condition 
for the formation of a double-torus 
in nebulae fed  by a pulsar with a high magnetic inclination and 
a weakly magnetized wind. Such nebulae are 
distinguished by the presence of a  barely magnetized and overcompressed 
equatorial flow with pulsations -- oblique and normal 
shocks --  by means of which the flow equalizes its pressure with 
neighboring regions of the nebula. Our current simulations confirm 
that the equatorial flow exhibits pulsations in the 3D case as well.
They can be seen in the pressure maps in Fig.\ \ref{3D_temp_evol_B_2}, 
describing  the full-dimensional  \textit{X3}\ model  (in Table 
\ref{Table_2D_3D}). This model, like the Vela model in 
\cite{Ponomaryov+21}, is characterized by a high inclination, 
weakly magnetized wind, and expansion strongly constrained by the 
external flow.

Finally, the strong external flow appears to be necessary 
for spectral modeling of Vela.
Its hard X-ray photon indices can be understood 
in a model in which a pulsar wind collides with a slightly supersonic 
flow. This was  shown in Monte Carlo simulations of particle 
acceleration in  collision of such flows \citep{BSPWN_2017}.
Interestingly, this model predicts a significant flux of synchrotron 
radiation from Vela in the MeV range. This is shown in the right panel 
on Fig.\ \ref{MC_spectra} which depicts the simulated spatially-resolved 
synchrotron spectra of Vela.
They were obtained by integrating, along certain lines of sight, the 
synthetic synchrotron emissivity calculated in each cell of a 
three-dimensional grid covering the volume of the model nebula.  
The particle spectrum in each cell was simulated using Monte Carlo approach.
The color of the synchrotron spectra in Fig.\ \ref{MC_spectra} indicates where 
the corresponding line of sight intersects the nebula: \textit{red}  -- 
near its wind termination shock; \textit{blue} -- near its bow shock, formed 
by the interaction with the external flow; \textit{green} --
near the contact discontinuity between these two shocks.     
Note that the model predicts the MeV flux only a few times lower 
than the X-ray flux detected by Chandra.

The potentially detectable MeV flux is likely a common feature of 
astrospheres  of pulsars -- both isolated and in gamma-ray 
binaries --  that are  energetic enough to produce charged particles 
in the PeV-range. In Vela and Crab, the pulsars can accelerate sub-PeV 
particles due to their high spin-down luminosity --- $\dot{E} \sim 7 
\times 10^{36}$ and $5 \times 10^{38} \ergs$  (almost completely 
converted into the energy of the pulsar wind), while pulsars in gamma-ray 
binaries  --- even with  a lower spin-down luminosity --- due to the 
strong magnetic field in the powerful stellar wind of their massive companion.

\section{Conclusions}
\label{sec:concusion}
In this paper, we considered modeling the structure of the wind collision region 
in gamma-ray binaries, where the relativistic outflow of a compact object collides 
with the strongly magnetized wind of an early-type massive companion star.
The collision wind region can have a few a.u. scale and can't be resolved 
in observations.
To demonstrate the ability of colliding winds,
we  simulated  the interaction of a strongly magnetized stellar wind with a
relativistic pulsar wind, based on  well-established models of isolated pulsar wind nebulae. 
The  models setups were thoroughly tested using a wealth of high-resolution radio, 
optical, and X-ray images of extended isolated galactic  nebulae (such as  Crab and Vela) and 
their spatially resolved spectra.
We believe that the process of accelerating  ultra high energy particles  by intermittent 
relativistic turbulence in the collision region can work reasonably  well in  gamma-ray binaries  
containing black holes accreting at critical rates and producing powerful outflows.

The present study shows that two-dimensional (2D) and full-dimensional (3D) 
models produce very similar results when simulating relativistic MHD flows in 
gamma-ray binaries. This indicates that the planar 2D models as those 
considered by  \citet{Bykov+24} can be safely used to study gamma-ray 
binaries as potential  Pevatrons, particle (proton) accelerators up 
to petaelectronvolt (PeV) energies.  
Previously, the qualitative similaritiy of 2D and 3D models were demonstrated 
by \citet{BoschRamon+15} within the framework of the relativistic hydrodynamic 
approach when studying the influence of orbital motion on the pattern of flows
in gamma-ray binaries. At contrast to the HD approach, our simulations take 
into account  the strong Gauss-range magnetic field, that may be present 
in the wind of a massive companion (an early-type star) at distances up to 
a few astronomical units from it. The orbital motion of a compact companion 
(a pulsar) may bring it to these distances, especially in short-period or 
highly eccentric gamma-ray binaries. 

The strong stellar-wind field causes the nebula to appear elongated, 
squeezing it across the field. Similar elongated shapes can be 
acquired by plerionic-type supernova remnants in  magnetized circumstellar 
environments, as shown by non-relativistic 2D MHD simulations of \citet{Meyer+24}.
At the same time, the equatorial size of the nebula remains virtually unchanged 
even in a field $\sim 3$ G. The same applies to the size of the X-ray torus, 
since this synchrotron feature of the nebula is formed precisely by 
its equatorial flows. The torus can  acquire a tilt toward the nebula's  equatorial plane 
if the magnetic field of the stellar wind makes an angle with 
the rotation axis of the pulsar. 
The tilt can change gradually along the pulsar’s orbit, reaching 
a minimum at apastron and a maximum at periastron.

A strong field is of paramount importance 
for accelerating particles, as it can ensure their confinement 
in the accelerator, the region where the pulsar wind and the stellar wind 
collide. This compact region encompasses the pulsar wind nebula and the
dense  magnetic cocoon that formed around it from the stellar wind material.
We found that the field in the 2D and 3D cocoons can reach a few Gauss, even 
though it is sub-Gaussian in the unperturbed stellar wind. 
The  cocoon is capable of confining  PeV  particles within  
the nebula for only a few hours, but this is enough for them to gain energies 
above 10 PeV before escaping the wind collision region.
Thus, the Pevatron accelerator in gamma-ray binaries 
can 
be switched on for a quite short interval of time during each orbital period. 
This does not prevent it from accelerating protons well above
the limit determined by the magnetospheric potential of the pulsar. 
(In principle, the acceleration can be continued if PeV protons 
emerging from the accelerator can be upscattered by magnetic 
inhomogeneities in the stellar wind).

Inside the nebula, the rapid acceleration can be provided by fast flows 
capable of producing  highly-magnetized, relativistic  clumps with scales 
comparable to or larger than  the gyroradii of PeV protons. Due to their high 
Lorentz factors, the clumps can increase  the energy of sub-PeV particles 
above PeV in just a few successful scatterings. We show that such clumps 
can arise in 2D and 3D nebulae, although they seem to form differently 
from  usual magnetic inhomogeneities in the flows. The maximum Lorentz 
factors of the clumps depend on the numerical resolution of the model, 
which is typical situation in an ideal magnetohydrodynamics, in which 
the real viscosity is substituted by a numerical one determined by the minimum 
spatial (and temporal) scales resolved by the numerical grid. 
In our higher resolution 2D models, the Lorentz factor can reach 
4.5. In computationally expensive
3D models with fewer mesh nodes per  nebula volume, the factors are somewhat lower. 
It remains to be seen whether this is due to their lower resolution or to a smoother 
pattern of flows. Higher resolution simulations are required for this study. 
Although direct numerical simulations  in \cite{Bykov+24} clearly showed
the ability of gamma-ray binary models to accelerate
multi-PeV protons, the limited spatial resolution of these models
prevented us from obtaining proton spectra over the wide energy range necessary for interpreting the observations.
Therefore, we simulated these spectra using test-particle Monte Carlo simulations, in which proton scattering rates 
in intermitten relativistic flows were parameterized over a wide energy range.
The spectra are shown in the left panel of Fig. \ref{MC_spectra}. Recent kinetic particle-in-cell simulations of particle acceleration by relativistic turbulence \citep{2025arXiv251215239L,2025PhRvD.112l3028L} demonstrated that the mechanism is promising indeed for explaining the origin of the highest-energy particles.  

In the model presented in this paper, we considered a powerful pulsar wind as the source 
of the relativistic outflow. In fact, many important features of the model hold for gamma-ray binaries 
where the compact companion is an accreting black hole. The powerful galactic microquasar Cygnus X-3  
is a  high-mass X-ray binary (HMXRB) with a period $\sim$ 4.8 hours;  its compact object -- either 
a black hole or a neutron star -- accretes matter from a massive Wolf-Rayet companion star. 
Cyg X-3 produces, apart from X-ray emission, jet-driven radio flares.
The LHAASO observatory detected time-modulated ultra-high-energy gamma-ray emission from this source, with energies up to a few PeV 
 \citep{2025arXiv251216638T}.
If the modulated PeV photon radiation is produced by the photomeson process \citep{2025arXiv251216638T}, 
it would require protons accelerated well above ten PeV. The proton spectra presented in Fig. \ref{MC_spectra} (left panel), obtained from Monte Carlo simulations, demonstrate that a relativistic outflow colliding with the highly magnetized wind of a massive star can accelerate particles to multi-PeV energies. The resulting specific two-peak spectra may be useful for interpreting LHAASO data.

\begin{acknowledgments}
 We acknowledge Georgy Ponomaryov for his contribution to our numerical studies of PWNe and gamma-ray binaries. Monte Carlo simulations by A.E.P. were supported by the Foundation for the Advancement of Theoretical Physics and Mathematics ``BASIS''. Data analysis by K.P.L.  was supported by the baseline project FFUG-2024-0002 at the Ioffe Institute.  Some of the modeling was performed at the ``Tornado'' subsystem of the St.~Petersburg Polytechnic University Supercomputing Center.
\end{acknowledgments}

\appendix

\section{rMHD modeling setups} \label{sec:Appenix-A-rMHD}
Our rMHD simulations of gamma-ray binaries are based on Relativistic MHD Module 
of the code PLUTO \citep{Mignone+07}. In this study, we assume that the binary 
system harbors a pulsar orbiting a massive early-type star (here, a Be star). 
We examine the local structure of the region where the pulsar wind and the stellar 
winds collide.  This region encompasses the pulsar wind nebula  and the dense cocoon   
that forms around it from the stellar wind material. 
The stellar wind is initialized as a homogeneous, strongly magnetized 
flow with co-directional magnetic field and velocity, following the 
arguments in Sect.\:\ref{sec:model}. The wind's parameters 
are given in Table \ref{Table_2D_3D}; they are chosen to illustrate the 
variety of conditions  that the pulsar might encounter where its orbit 
takes it close to the Be star. Our simulations indicate that  the 
local MHD structure of the pulsar wind nebula depends more than moderately 
on the velocity and density of the stellar wind, as long as the latter 
is strongly magnetized.

To inflate the nebula, we apply a simplified  model of the pulsar wind that 
is widely used in modern rMHD simulations of isolated nebulae \citep[e.g.,][]{delZanna+04,Porth+14,Olmi+16,BuhlerGiomi16,Ponomaryov+23}, in the formulation
given in \cite{BuhlerGiomi16}.
In this model, the wind is cold and its energy flux density 
is latitudinally dependent, with a maximum at the plane of the rotational 
equator of the pulsar (this plane determines the equatorial plane of the nebula):
\begin{equation}
    f_{tot\, }(r, \theta) = \frac{\dot{E}}{L_{\, 0}} \frac{1}{r^{\, 2}}\cdot 
    \left( \sin^2 \theta + \varepsilon  \rule{0pt}{1.5ex} \right)\, .
\end{equation}
Here $r$ is a distance from the pulsar, $\theta$ -- colatitude, 
$L_{\,0} = 4\pi\:(2/3 + \varepsilon)$ -- normalization constant, 
$\varepsilon = 0.02$ prevents the flux from vanishing at the poles.
The energy flux is divided into magnetic and kinetic components,
$f_{tot} = f_m + f_k$, the ratio of which is determined by the 
magnetization $\sigma = f_m/f_k$ of the cold pulsar wind:
\begin{equation}
\begin{aligned}
    & f_{m\,}(r, \theta) = \frac{\sigma(\theta)\cdot f_{tot\:}(r, \theta)}{1 + \sigma(\theta)}; \\
    & f_{k\,}(r, \theta)  = \frac{f_{tot\:}(r, \theta)}{1 + \sigma(\theta)}; \\
    & \sigma(\theta) = \frac{\tilde{\sigma}(\theta)\cdot \chi_{\alpha} (\theta)}{1 +  \tilde{\sigma}(\theta)\cdot 
    (1 - \chi_{\alpha} (\theta)\:)}
    \label{eq:fluxes}
\end{aligned}
\end{equation}

Magnetic field of the wind is purely  toroidal and frozen into the plasma.
The wind magnetization is latitudinally dependent: it almost vanishes 
near the rotational equator and toward the poles, and has a maximum 
at the middle latitudes. These features are accounted with the functions 
$\tilde{\sigma}$ and $\chi$:
\begin{equation}
\begin{aligned}
    & \tilde{\sigma}(\theta) = \sigma_0 \cdot {\rm min} \left\{1\, ,\;\, 
    \frac{\theta^2}{\theta_0^2}\, , 
    \;\,     \frac{(\pi - \theta)^2}{\theta_0^2}  \rule{0pt}{1.ex}\right\}; \\
    & \chi_{\alpha}(\theta) = 
    \begin{cases} 
    (2 \phi_a (\theta)/\pi - 1)^2 & \mbox{if}\;|\pi/2 - \theta| < \alpha  \\
        1 & \mbox{otherwise}
    \end{cases}
    \label{eq:tilde-sigma-chi}
\end{aligned}
\end{equation}
with $\phi_{\alpha} (\theta) = \arccos{(-\cot(\theta) \cot(\alpha))}\:$ and 
a typical value of $\theta_0 = 10^{\circ}$. Here $\sigma_0$ is the initial 
magnetization, and the function $\tilde{\sigma}$ describes its decrease toward 
the poles due to the azimuthal nature of the field $(B\equiv B_\varphi \propto 
\sin\theta)$. The function $\chi_{\alpha}$ accounts for the fact that the wind 
magnetic field alternates at the equatorial latitudes within the angular sector 
$\pi/2 \pm \alpha$. As is known, the field of the pulsar changes sign at the 
pulsar's magnetic equator. The latter usually  makes a certain angle $\alpha$ 
(called \textit{magnetic inclination}) to the rotational equator of the pulsar. 
As the pulsar rotates, the wind magnetic field changes sign at the equatorial 
latitudes every half-period of rotation and hence the wind becomes striped.
It carries the stripes of plasma of alternating magnetic polarity which fill 
the angular sector $\pi/2\pm \alpha$. Either on their way to the wind termination 
shock, or due to their compression right at the shock, the stripes can annihilate 
\citep{Lyubarsky+03,Komissarov13, Cerutti+20}, so the magnetic field in 
the equatorial sector $\pi/2\pm \alpha$ dissipates almost completely.
For the pulsar wind's density, magnetic field and velocity we have:
\begin{equation}
\begin{aligned}
    & \rho_{pw\,}(r, \theta) = f_{k\,} (r, \theta) / \varGamma_{pw}^{\,2} \,c^3\:,\\
    & B_{pw\,}(r, \theta) = \pm \sqrt{\:4 \pi f_{m\, } (r, \theta) / c \rule{0pt}{1.8ex}}\:, \\
    & u_{pw}/c  = \sqrt{\:1 - \varGamma_{pw}^{\, -2}\:} .
\end{aligned}
\end{equation}
The wind is radially directed  and has a Lorentz-factor $\varGamma_{pw} \gg 1$.
To maintain numerical stability,  $\varGamma_{pw} =10$ is usually assumed, 
although this choice does not affect the calculation results as far as 
$\varGamma_{pw} \gg 1$ (see \cite{BuhlerGiomi16} for details).
Our numerical 2D/3D grids  are Cartesian, with the y-axis aligned with 
the pulsar's rotation  and no $z$-axis in 2D space. The coordinates of 
the Cartesian and spherical systems are related as 
$r=(x^2 + y^2 + z^2)^{1/2}$ and $\theta = \arccos^{\,}(y/r)$.
\begin{table*}
\centering 
\footnotesize
\begin{tabular}{lccccccccccccccc}\hline
    run & $B_{sw}$ & $\psi, {}^{\circ}$ & $n_{sw}$ & $p_{sw}$ & $\dot{E}$ & $\alpha, {}^{\circ}$ & $\sigma_0$ & $\varGamma_{pw}$ & box, au$^2$/au$^3$ & core, au$^2$/au$^3$ & $r_{in}$, au & $N_x$ & $N_y$; $N_z$ & $\Delta$, au \\ \hline
  A1 & 0.5 & 75 & $3 \times 10^4$ & $10^{-7}$ & $10^{37}$ & 45 & 3 & 100 & $40\times40$ & $40 \times 40$ & 0.15 & 1600 & 1600  & 0.025\\
  A2 & 1 & 75 & $3 \times 10^4$ & $10^{-7}$ & $10^{37}$ & 45 & 3 & 100 & $40\times40$ & $40 \times 40$ & 0.15 & 1600 & 1600  & 0.025\\
  A3 & 2 & 75 & $3 \times 10^4$ & $10^{-7}$ & $10^{37}$ & 45 & 3 & 100 & $40\times40$ & $40 \times 40$ & 0.15 & 1600 & 1600  & 0.025\\
  A4 & 3 & 75 & $3 \times 10^4$ & $10^{-7}$ & $10^{37}$ & 45 & 3 & 100 & $40\times40$ & $40 \times 40$ & 0.15 & 1600 & 1600  & 0.025\\
  A5 & 3 & 45 & $3 \times 10^4$ & $10^{-7}$ & $10^{37}$ & 45 & 0.3 & 100 & $40\times40$ & $40 \times 40$ & 0.15 & 1600 & 1600  & 0.025\\
  A6 & 3 & 45 & $3 \times 10^4$ & $10^{-7}$ & $10^{37}$ & 80 & 3 & 100 & $40\times40$ & $40 \times 40$ & 0.15 & 1600 & 1600  & 0.025\\
    X1 & 0.5 & 75 & $2 \times 10^5$ & $10^{-4}$ & $10^{35}$ & 80 & 0.3 & 4 & $2\times1.46\times1.46$ & $0.6 \times 0.6 \times 0.6$ & 0.02 & 340$+$600$+$340 & 252$+$600$+$252 &  0.001\\
     X2 & 1 & 75 & $2 \times 10^5$ & $10^{-4}$ & $10^{35}$ & 80 & 0.3 & 4 & $2\times1.46\times1.46$ & $0.6 \times 0.6 \times 0.6$ & 0.02 & 340$+$600$+$340 & 252$+$600$+$252 &  0.001\\
      X3 & 2 & 75 & $2 \times 10^5$ & $10^{-4}$ & $10^{35}$ & 80 & 0.3 & 4 & $2\times1.46\times1.46$ & $0.6 \times 0.6 \times 0.6$ & 0.02 & 340$+$600$+$340 & 252$+$600$+$252 &  0.001\\ 
       X4 & 3 & 75 & $2 \times 10^5$ & $10^{-4}$ & $10^{35}$ & 80 & 0.3 & 4 & $2\times1.46\times1.46$ & $0.6 \times 0.6 \times 0.6$ & 0.02 & 340$+$600$+$340 & 252$+$600$+$252 &  0.001\\
       X5 & 0.5 & 45 & $2 \times 10^6$ & $10^{-3}$ & $10^{37}$ & 80 & 0.3 & 4 & $3\times3\times3$ & $0.6 \times 0.6 \times 0.6$ & 0.027 & 363$+$451$+$363 & 363$+$451$+$363 &  0.0013\\
       X6 & 0.5 & 75 & $2 \times 10^5$ & $2 \times 10^{-4}$ & $10^{35}$ & 60 & 0.1 & 4 & $2\times2\times2$ & $0.6 \times 0.6 \times 0.6$ & 0.03 & 240$+$400$+$240 & 240$+$400$+$240 &  0.0015\\ \hline
\end{tabular}
\normalsize
\caption{Nebula models parameters.  
\underline{For the stellar wind \textit{(sw):}}
    $\bm{B}_{sw}$  is the magnetic field in Gauss; $\psi$ is the angle between 
    $\bm{B}_{sw}$ and pulsar’s spin axis; $n_{sw}$ and $p_{sw}$ are the number 
    density and pressure, in cgs units. In all setups, the stellar wind velocity
    is taken to be $300\; \mbox{km}\,\mbox{s}^{-1}$, as seen in the pulsar frame 
    of reference \citep[see, e.g.][]{Bogovalov+08}. \underline{For the pulsar wind \textit{(pw):}}\ 
    $\varGamma_{pw}$ and $\sigma_0$ -- the Lorentz-factor and initial magnetization; 
    $\dot{E}$ -- the total power (the pulsar’s spindown luminosity, in erg\,s$^{-1}$); 
    and $\alpha$ -- the angle between the rotational and magnetic axes of the pulsar.
\underline{For the modeling domain:}  
    the ``box'' is its full spatial dimension, and the ``core'' is its central region  
    ($x_1 < x < x_2$, $\;y_1 < y < y_2$, $\;z_1     < z < z_2$)
    with a uniform grid of resolution $\Delta$ and with a small area of radius 
    $r_{in}$ in the center, from which the pulsar  wind is continuously injected. 
    $N_x$, $N_y$, $N_z$ are the total numbers of the grid nodes in the ``box'' in 
    the $x$, $y$, $z$  directions, respectively.  In 2D setups the $z$ axis is 
    absent, while in 3D setups it has the same subdivision as the y-axis
    $(N_y=N_z)$.
In 2D models \textsl{A1}--\textsl{A6}, the entire grid is uniform
(``core''$\equiv$``box''),  and $N_x=N_y$. 
In 3D models \textsl{X1}--\textsl{X6}, the grid is patchy; its x-axis 
is subdivided as follows: its step increases logarithmically with  distance
below $x_1$ and above $x_2$, and is uniform in the ``core'' 
between $x_1$ and $x_2$;  the number of nodes in each interval of $x$ is 
represented by the corresponding term in the sum in the $N_x$ column. 
The same subdivision is applied to the $y$ and $z$ axes. 
}
\label{Table_2D_3D}
\end{table*}

\section{Simplified anisotropic Monte-Carlo model}\label{sec:Appendix-B-anis-MC}
To simulate particle acceleration in the wind collision zone of a gamma-ray binary -- accounting for efficient particle scattering by turbulence with a wide range of spatial scales -- we developed a massively parallelized  
numerical code based on the Monte Carlo approach.
The code 
is three-dimensional and 
takes into account the  propagation of 
charged particles in a turbulent magnetic field with a strong regular component. Based on the simplified model of flow structure, it allows us to study the acceleration in the collision zone between a strongly magnetized stellar wind and a relativistic outflow --produced either by a pulsar (as in the present study) or by a black hole. The description below assumes a gamma-ray binary with a pulsar.

The computation domain is a cylinder with radius $R_{box}$ and length $l_{box} 
= 2R_{box}$ and a uniform magnetic field $B_{box}$ parallel to the cylinder's 
axis. Two colliding winds -- pulsar and stellar -- are modeled as two  
magnetized uniform flows, each of which occupies half of the cylinder. The winds' 
velocities --  $\bm{u}_{pw}$ and $\bm{u}_{sw}$ -- are counter-directed and 
parallel to the cylinder's axis.  The contact discontinuity (CD) between the winds 
is represented by a narrow buffer region with zero flow velocity. 
A small cylindrical subregion emdedded in the region of the pulsar wind models the relativistic clump with $\varGamma > 2$. 
The simulation parameters reproduce the conditions in  gamma-ray binaries:
$R_{box} \approx 3\:$AU, $\;B_{box} = 0.1\:$G, $\;u_{pw} = 0.57\, c$ and 
$\;u_{sw} = 800\:$\kms.  The clump has a size of $0.6 \times 1.2$ AU
and a flow velocity with a high Lorentz factor $\varGamma$. We considered the cases 
with $\varGamma=\:$3, 4.5 and 6. The result of the Monte Carlo simulation 
is shown in Fig.\ \ref{MC_spectra}, on the left panel. 

The simulation of acceleration begins by injecting a population of nonthermal 
particles into the CD region. The injected population is assumed to be 
pre-accelerated (e.g., on the wind termination shock, TS), having an 
initial spectrum $f_{TS}\left(E\right) \propto E^{-2.2}$.
The particles then move along their (helical) trajectories 
prescribed by Newtonian laws for a certain free-flight time $t_{mfp}$. After that, 
each particle is scattered isotropically (in the rest frame of a local background 
flow), and moves again for time $t_{mfp}$, and so on, until it leaves the computational 
domain through its border with the free escape boundary conditions.
The free-flight time $t_{mfp}$ is determined by the mean free path  
$\lambda_{mfp} = c\cdot t_{mfp} = \eta R_{g}$, where $R_g$ is the particle gyroradius, with $\eta = 6$ in the pulsar wind's and the stellar wind's regions and $\eta = 12$ in the CD region.
%

\end{document}